\documentclass[10pt, journal,fleqn]{IEEEtran}  

\IEEEoverridecommandlockouts     

\usepackage{color,amsmath,amstext,amsfonts,amssymb, mathrsfs,mathtools}
\usepackage{graphicx,algorithm,algorithmic}
\usepackage{tikz}
\usepackage{setspace}
\usepackage{array} 
\usepackage{booktabs}  
\usepackage{bbm}
\usepackage{wasysym}
\usepackage{textcomp}
\usepackage{hyperref} 
\usepackage[sort,compress]{cite}

\newtheorem{lemma}{Lemma}
\newtheorem{assumption}{Assumption}

\newtheorem{theorem}{Theorem}
\newtheorem{definition}{Definition}

\renewcommand{\thesubsubsection}{{\it \Alph{subsection}.\arabic{subsubsection}.}}

\newcounter{l1}
\newcounter{l2}
\newcounter{l3}
\newcommand{\bdotlist}{\begin{list}{$\bullet$}{}}
\newcommand{\bboxlist}{\begin{list}{$\Box$}{}}
\newcommand{\bbboxlist}{\begin{list}{\raisebox{.005in}{{\tiny $\blacksquare$ \ \ }}}{}}
\newcommand{\bdashlist}{\begin{list}{$-$}{} }
\newcommand{\blist}{\begin{list}{}{} }
\newcommand{\barablist}{\begin{list}{\arabic{l1}}{\usecounter{l1}}}
\newcommand{\balphlist}{\begin{list}{(\alph{l2})}{\usecounter{l2}}}
\newcommand{\bAlphlist}{\begin{list}{\Alph{l2}.}{\usecounter{l2}}}
\newcommand{\bdiamlist}{\begin{list}{$\diamond$}{}}
\newcommand{\bromalist}{\begin{list}{(\roman{l3})}{\usecounter{l3}}}

\providecommand{\norm}[1]{\lVert#1\rVert}

\newcommand{\beq}{\begin{equation}}
\newcommand{\eeq}{\end{equation}}

\title{\LARGE \bf
Memory Augmented Neural Network Adaptive Controllers: Performance and Stability
}

\author{Deepan Muthirayan$^a$ and Pramod P. Khargonekar $^a$  
\thanks{$^a$ Electrical Engineering and Computer Sciences, University of California, Irvine, CA 92697. Email: \{dmuthira, pramod.khargonekar\}@uci.edu}
\thanks{Supported in part by the National Science Foundation under Grant Number ECCS-1839429.}
}

\begin{document}
\maketitle
\markboth{Submission for IEEE Transactions on Automatic Control}{Deepan \emph{et al.}} 

\thispagestyle{empty}
\pagestyle{empty}

\begin{abstract}
In this paper, we propose a novel control architecture, inspired from neuroscience, for adaptive control of continuous-time systems. 
The proposed architecture, in the setting of standard Neural Network (NN) based adaptive control, augments an {\it external working memory} to the NN. The controller, through a write operation, writes the hidden layer feature vector of the NN to the external working memory and can also update this information with the observed error in the output. Through a read operation, the controller retrieves information from the working memory to modify the final control signal. 

%

First, we consider a simpler estimation problem to theoretically study the effect of an external memory and prove that the estimation accuracy can be improved by incorporating memory. We then consider a model reference NN adaptive controller for linear systems with matched uncertainty to implement and illustrate our ideas. We prove that the resulting controller leads to a {\it Uniformly Bounded} (UB) stable closed loop system. Through extensive simulations and specific metrics, such as peak deviation and settling time, we show that memory augmentation improves learning significantly. 
Importantly, we also provide evidence for and insights on the mechanism by which this specific memory augmentation improves learning.
\end{abstract}

\section{Introduction}
\label{sec:int}
A human's learning system is arguably the most versatile and flexible learning system known so far. For example, humans excel at many tasks such as concept learning, scene understanding, and language understanding, where the capabilities of machines are still found lacking \cite{lake2017building}. While machine learning (deep learning, reinforcement learning) algorithms have been able to match human-level performance in many tasks \cite{krizhevsky2012imagenet, hinton2012deep, mnih2015human}, they still lack ``human-like'' learning capabilities in many aspects. 
For example, the current generation of machine learning algorithms typically need large datasets, whereas humans can learn from just a few examples. Humans can also adapt to completely unseen environments, something that learning machines find challenging. We believe that these aspects  are potentially highly relevant from a control perspective. Thus, it is natural to draw inspiration from, and take advantage of, knowledge in neuroscience and cognitive science to achieve challenging goals such as adaptation, flexibility, and autonomy in engineered systems. 

Memory plays a central role in learning and cognition tasks in humans \cite{tulving1985many,roediger2017typology,gershman2017reinforcement}. 
Inspired by these insights in neuroscience and cognitive science, we focus on the following questions: can control algorithms improve their learning and performance by incorporating memory structures inspired from human-like memory systems? If so what is the architecture and the learning algorithm? Is the architecture universal or problem dependent? And does it improve learning in all scenarios? These are hard questions, and to the best of our knowledge, have been relatively under-explored. 
We note that, from a traditional dynamic systems and control perspective, the state of the nonlinear dynamic controller constitutes the ``memory''. {\em However, we are proposing specific memory modules that will augment the state of the dynamic nonlinear controller and potentially lead to new learning and control capabilities.}

As an initial step towards these longer-term goals, we have chosen a well-studied Neural Network (NN) adaptive control setting. We note that the literature on NN based adaptive control is very extensive, see for example, \cite{narendra1990identification, yecsildirek1995feedback, lewis1996multilayer, narendra1997adaptive, kwan2000robust, calise2001adaptive, chen2001nonlinear, ge2004adaptive} and the references cited there. The setting is the standard adaptive control setting where the unknown nonlinear function is approximated by a neural network. In addition, we consider nonlinear uncertainties that can vary with time. 

We propose a novel control architecture that {\it augments an external working memory to the regular NN adaptive controller}. It is certainly conceivable that the speed of learning of an adaptive and learning controller can be improved by increasing the learning rate \cite{stepanyan2012adaptive, gibson2012adaptive, yucelen2014improving}. By contrast, in the architecture we propose, the {\it learning is improved by the use of information in the external memory}. In machine learning, the idea that NNs with additional external memory have advantages in requiring less data, goes back to early 2000's, and possibly before. In the work by Hochreiter et. al \cite{hochreiter2001learning}, the authors showed that, LSTMs (Long Short Term Memory) which have an inherent memory can quickly learn never-before-seen quadratic functions with a low number of data samples. In deep learning, architectures with external working memory were proposed recently in \cite{graves2014neural}, \cite{santoro2016meta, parisotto2017neural}. It was demonstrated that the addition of an external working memory to these models improved their performance. Our contribution is to leverage this insight in proposing a novel NN based control architecture that includes an external memory. We propose a specific external memory design to augment the NN adaptive controller.

In Section \ref{sec:est}, we analyze a general external memory for augmenting the estimation of a signal that is an output of an unknown function and prove that the estimation error can be reduced by memory augmentation. In Section \ref{sec:arch} we propose the {\it Memory Augmented Neural Network} (MANN) adaptive controller. 
 In Section \ref{sec:mem-int} we discuss the working memory for the NN adaptive controller and establish dynamic stability of the closed loop system. 
 Finally, in Section \ref{sec:sim-results} we provide a detailed set of simulation results and discussion substantiating the improvements in learning obtained by memory augmentation for couple of applications. 

\section{Memory Augmented Control Architecture}
\label{sec:arch}

\begin{figure}
\center
\begin{tikzpicture}[scale = 0.45]

\draw [draw=blue, fill=blue, fill opacity = 0.1, rounded corners, thick] (-2, 4) rectangle (2, 5);
\draw (0, 4.5) node[align=center] {\tiny $f$};

\draw [draw=blue, fill=blue, fill opacity = 0.1, rounded corners, thick] (-2, 0) rectangle (2, -1);
\draw (0, -0.5) node[align=center] {\tiny Controller};
\draw  [<->, thick] (0,-1) -- (0,-2);

\draw [draw=blue, fill=blue, fill opacity = 0.1, rounded corners, thick] (-2, -2) rectangle (2, -3);
\draw (0, -2.5) node[align=center] {\tiny Working Memory};

\draw [draw=blue, fill=blue, fill opacity = 0.1, rounded corners, thick] (-2, 1) rectangle (2, 3);
\draw (0, 2) node[align=center] {\tiny Plant};
\draw  [->, thick] (-2,-0.5) -- (-3,-0.5) -- (-3,1.5) -- (-2,1.5);
\draw  [->, thick] (2,1.5) -- (3,1.5) -- (3,-0.5) -- (2,-0.5);
\draw  [->, thick] (-2,4.5) -- (-3,4.5) -- (-3,2.5) -- (-2,2.5);
\draw  [->, thick] (2,2.5) -- (3,2.5) -- (3,4.5) -- (2,4.5) ;
\draw (-3.25, 1.5) node[align=center] {\small $u$};
\draw (-3.25, 2.5) node[align=center] {\small $w$};
\draw (3.25, 1.5) node[align=center] {\small $x$};
\draw (3.25, 2.5) node[align=center] {\small $z$};
\draw (-3.25, 2) node[align=center] (d) {\small $s$};
\draw  [->, thick] (d) -- (-2, 2);
\draw (3.25, 2) node[align=center] (e) {\small $e$};
\draw  [<-, thick] (e) -- (2, 2);

\draw [draw=blue, fill=blue, fill opacity = 0.1, rounded corners, thick] (5, 0.25) rectangle (9.25, 4);
\draw [draw=blue, fill=blue, fill opacity = 0.1, rounded corners, thick] (7.75, 1) rectangle (9, 3);
\draw (8.4,2) node[align=center] {\tiny NN};
\draw (6,2) node[circle , draw, thick] (b) {\tiny $+$};
\draw [->, thick] (b) -- (4.5,2);
\draw [->, thick] (7.75,2) -- (b);
\draw [->, thick] (6,3.5) -- (b);
\draw [->, thick] (6,0.5) -- (b);
\draw (7, 4.25) node[align=center] {\tiny Neural Adaptive Controller};
\draw (4.25,2) node[align=center] {\small $u$};
\draw (6,3.75) node[align=center] {\small $v$};
\draw (6.5,0.75) node[align=center] {\small $u_{bl}$};
\draw (7.3,2.25) node[align=center] {\small $u_{ad}$};
\draw (10.5,2.25) node[align=center] (c) {\small $x$};
\draw [->,thick] (c) -- (9.25,2.25);

\draw [draw=blue, fill=blue, fill opacity = 0.1, rounded corners, thick] (5, -2) rectangle (9.25, -3);
\draw (7, -2.5) node[align=center] {\tiny Working Memory};
\draw  [->, thick] (6.5,-2) -- (6,0.25);
\draw  [->, thick]  (8.25,0.25) -- (7.75,-2);
\draw (5.75,-1) node[align=left] {\tiny Read};
\draw (8.5,-1) node[align=right] {\tiny Write};

\end{tikzpicture}
\caption{Left: general controller augmented with working memory. Right: memory augmented neural adaptive controller. The function $f$ is the uncertainty. The signal $s$ is the command or reference signal.}
\label{fig:cwmem}
\end{figure}
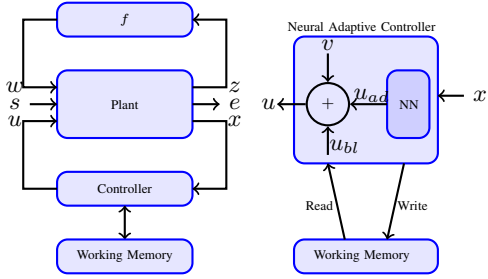

Our envisioned general architecture augments an external working memory to the general dynamic feedback NN adaptive controller, as depicted in Fig. \ref{fig:cwmem}. There are potentially numerous ways to formulate concrete algorithms based on this general architecture. In this paper, we specialize it to the specific controller that augments an external working memory to a neural network, which represents an implicit memory. The intuitive idea is to leverage the combination of an external working memory, which can store relevant information for retrieval and later use, and the implicit memory in the NN, to achieve better learning in estimation and control.


In a typical neural adaptive control setting, the control law computes the control input $u$ to the plant based on the state $x$ and error feedback $e$. The control input is a combination of base controller $u_{bl}$, a NN output term $u_{ad}$ and a ``robustifying term'' $v$ \cite{lewis1996multilayer, lewis1998neural}. The robustifying term is needed to provide some level of robustness to the closed loop system. Thus, the final control input is given by
\beq u = u_{bl} + u_{ad} + v. \label{eq:claw} \eeq 

In our proposed control law, {\it  the NN output $u_{ad}$ is modified by the information read from the external working memory.} More specifically, we propose to modify the hidden layer output of the NN with the information read from the external working memory. This design of memory augmentation is one of our contributions.


In machine learning, Graves et al. \cite{graves2014neural} introduced the idea of augmenting an external working memory to a NN model, the Neural Turing Machine (NTM). 
It was demonstrated that the addition of their external working memory to a deep NN such as Long Short Term Memory (LSTMs) improved their performance. 

The controller in NTM has two memory operations (i) {\it Memory Write} and (ii) {\it Memory Read}. The write operation generates the memory content while the read operation retrieves useful information from the memory. The memory stores {\it \{key, value\}} pairs. We denote a key-value pair by $\{k_i, \mu_i\}$, where $i$ takes values in some (finite) set. The keys \{$k_i$\} contain information that is used by the Memory Read to determine the values to retrieve from the set \{$\mu_i$\}. Depending on the implementation, either (i) the key and the value can be the same and (ii) the key can be the location. The Read operation generates a query $q$, which is then matched with the keys to determine the values to read from the memory. The value that is read from the memory is used by NTMs to produce their final output.


The Memory Write equation in NTM has a {\it forget term} and an {\it update term}. The forget term in the Write operation allows the controller to gradually remove the contents in the memory that are irrelevant. On the other hand, the update term allows the controller to update the contents of the memory with the new information, provided it is relevant. This allows NTMs to retain information and also update with the new information, depending on their relevance. 

The external memory model we propose for the NN adaptive controller is inspired from these ideas. Our memory architecture, depicted in Fig. \ref{fig:cwmem}, is similar to NTM. It has an external memory with a Memory Read and Memory Write operation. The Memory Read is similar to the NTM. The Memory Write has a forget term and an update term like the NTM, and an additional third term that allows the memory to be updated with the observed error in the output. The additional update term allows the memory to be updated when the information from the regular update (the first update term) as indicated by the observed error is less accurate. 


 {\it Notation}: We denote the system state by, $x \in \mathrm{R}^n$, and the command signal by $s$. We consider a two-layer neural network for estimating the unknown function, $f(x)$, in the system dynamics. 
The estimated NN weight matrices are denoted by $\hat{W}, \hat{V}, \hat{b}_w$ and $\hat{b}_v$. Thus, the estimator of $f(x)$ is $\hat{f} = \hat{W}^T\sigma(\hat{V}^Tx + \hat{b}_v) + \hat{b}_w$, where the function $\sigma(\cdot)$ is the sigmoid function.

We introduce two vector signals, $\hat{\sigma}$ and $\hat{\sigma}'$. These signals appear in the NN update laws, as we shall see later. These signals are given by,
\[ \hat{\sigma}  = \left[ \begin{array} {c} \sigma(\hat{V}^Tx + \hat{b}_v) \\ 1 \end{array} \right], \nonumber \]
\beq \hat{\sigma}' = \left[ \begin{array} {c} \text{diag}(\sigma(\hat{V}^Tx + \hat{b}_v)\odot(1 - \sigma(\hat{V}^Tx + \hat{b}_v))) \\ \mathbf{0}^T \end{array} \right]. \nonumber \eeq
where $\mathbf{0}$ is a zero vector of dimension equal to the number of hidden layer neurons, $\odot$ is the Hadamard product. We say that a NN is irreducible if there are no compensating layer pairs (i.e. whose summation is not a constant) and no constant hidden layers. When we say that a function abruptly changes by $f \rightarrow g \ \text{at} \ t$, we imply that the function jumps instantaneously from $f$ to $g$ at time $t$. Later, we use the function $O(\cdot)$. When a scalar variable $\beta = O(\delta)$, it implies that $\exists$ a constant $c \ \text{s.t.} \ \vert\beta\vert \leq c \vert\delta\vert$. We denote the 2-norm by $\norm{\cdot}$ and the frobenious norm by $\norm{\cdot}_F$. We denote the space of non-negative real numbers by $\mathbb{R}_{\geq 0}$. We denote the Banach space of all continuous functions over a compact set $X$ with norm $\norm{f}_{C(X)} = \max_{x \in X} f(x)$ by $C(X)$. We denote the cartesian product of two sets $X$ and $Y$ by $X\times Y$. Given two vectors $a$ and $b$, let $[a;b]$ denote the stack of the two vectors with $a$ above $b$. We define the minimum and maximum eigenvalue of a matrix $X$ by $\lambda_{min}(X)$ and $\lambda_{max}(X)$ respectively.  

\section{External Working Memory Augmentation for Estimation}
\label{sec:est}

In this section, we demonstrate how external memory can improve the estimation of a signal that is an output of an unknown function. An estimation problem is simpler than the closed loop control problem and allows us to analyze the performance improvement mathematically. We demonstrate the benefit of memory by a comparative analysis of the estimator with memory and one without memory. We use a two timescale dynamical system framework for mathematical analysis of the estimators.

Consider a signal $y \in \mathbb{R}$, which is the output of an unknown time varying function $f_t$ with input signal $x \in \mathcal{K}$, a compact subset of $\mathbb{R}^n$:
\begin{equation}
y = f_t(x). 
\label{eq:estsys}
\end{equation} 
Consider a two layer neural network approximation $f_{t,a}$ of $f_t$ with weights $V_t \in \mathbb{R}^{n \times N}$, bias $b_t \in \mathbb{R}^{N \times 1}$ and $W \in \mathbb{R}^N$ and define the error in approximation $\delta_t$ as follows:
\begin{equation}
f_{t,a}(\tilde{x}) = W^T\sigma(V^T_tx+b_t), \ \text{s.t.} \ f_t(\tilde{x}) - f_{t, a}(\tilde{x}) =: \delta_t
\label{eq:estsys}
\end{equation}
Let $\tilde{x}^T= [x^T,~ 1]$ and $S^T_t = [V^T_t, b_t]$. Then, we can write $f_{t,a}(\tilde{x})$ compactly as
\begin{equation} 
f_{t,a}(\tilde{x}) = W^T\sigma(S^T_t\tilde{x})
\end{equation}

If the approximation error satisfies the error bound
\begin{equation}
\norm{\delta_t} \leq \overline{\delta}
\end{equation}
then the approximator $f_{t,a}(\tilde{x})$ is called a $\overline{\delta}$-approximator. 
We assume that the weight $W$ is constant, while $S_t$ and $\delta_t$ are time varying. The weights $S_t$ and $W$ are unknown and $N$ is known. 

The next lemma is aimed at elucidating the class of functions $f_t$ that admit such neural network approximators.  \begin{lemma}
For a finite $T >  0$, let $I := [0,T]$. Suppose that the map $F: I \rightarrow C(\mathcal{K}): F(t) := f_t$ is continuous. Then, for any $\overline{\delta} > 0$   there exist $N \in \mathbb{N}$, $W \in \mathbb{R}^{1 \times N}, S_t \in \mathbb{R}^{n+1\times N}$ such that for all $(t,x) \in I \times \mathcal{K}$,
\begin{equation}
\left\vert f_t(x) - W^T\sigma(S^T_t\tilde{x})\right\vert < \overline{\delta}
\end{equation}
\label{lem:app-capability-NNmodel}
\end{lemma} 
Thus, for the class of functions $f_t$ that vary continuously with respect to $t$, it is possible to find $\overline{\delta}$-approximators for any positive $\overline{\delta}$. Please see the Appendix for the proof. In proving the result we leverage the universal approximation theorem for continuous functions; see Theorem 2, \cite{hornik1991approximation}. 

Next, we define the estimators with and without memory below.

{\it Standard NN based estimator}: Let $\hat{W}$ and $\hat{S}$ denote the estimates of $W$ and $S_t$ respectively.  The estimate $\hat{y}$, and the error in estimation, $e$, are given by 
\begin{equation}
\hat{y} = \hat{W}^T\sigma(\hat{S}^T\tilde{x}), \ e = y -\hat{y}.
\label{eq:est-without-memory}
\end{equation}

{\it NN based estimator with external working memory}: For this estimator, the estimate, $\hat{y}_m$, and the error in estimation, $e_m$, are given by
\begin{equation}
\hat{y}_m = \hat{W}_m^T\left(\sigma(\hat{S}_m^T\tilde{x}) + \alpha M_r\right), \ e_{m} = y - \hat{y}_m,
\label{eq:update-NN-with-memory}
\end{equation}
where $\hat{W}_m$ and $\hat{S}_m$ are the estimates of $W$ and $S_t$ respectively, $M_r$ is the Memory Read output, and $\alpha \in (0, 1]$, is a constant. We denote the state of the memory by $\mu \in \mathbb{R}^{N \times n_s}$, where $n_s$ is the number of memory locations and the column vectors are the values of the respective locations. As described in Section \ref{sec:arch}, the memory is associated with a key $k_\mu \in \mathbb{R}^{n_k \times n_s}$. The Memory Read output is generated by the same process described in Section \ref{sec:arch}, and we functionally represent it by $M_r(\mu, k_\mu, \hat{S}_m, \tilde{x})$. Let $e_{m,a} = y_a - \hat{y}_m, y_{a} =  f_{t,a}(\tilde{x})$. The parameter $\alpha$ can be used to quantitatively tune the influence of the external working memory. 

Next, we introduce a definition on the equivalence of two NNs which we will use to define the initial condition of the dynamics corresponding to the two estimators.
\begin{definition}
{\it Two neural networks are input-output equivalent provided their input-output map $f_1(\cdot)$ and $f_2(\cdot)$ satisfy
$f_1(\tilde{x}) = f_2(\tilde{x})$ for all $\tilde{x}$.}
\end{definition} 

In the remainder of this section, we analyze the performance advantages of the memory based estimator. The performance advantage for the memory based estimator arises from the memory term $M_r$ that is added to the hidden layer of the estimator. This works particularly well when the variations in {$S_t$} are much faster than the variations in $\tilde{x}$. In the following, we model the variation of {$S_t, \delta_t$} and $\tilde{x}$ and the dynamics of the estimators as a two timescale system, where {$S_t, \delta_t$} can vary fast and $\tilde{x}$ varies slowly. The scaling $\epsilon ( \ll 1)$ controls the slow variation rate. The measurable function $\tilde{d}_f = (d_f, \delta_d): \mathbb{R}_{\geq 0} \rightarrow \mathcal{V}_f \subset \mathbb{R}^{n\times N}$ is the rate of change (or disturbance) of the hidden layer weight {$V_t$} and {$\delta_t$}, and the measurable function $d_s: \mathbb{R}_{\geq 0} \rightarrow \mathcal{V}_{s}$ models the rate of change of the input signal $\tilde{x}$. 

The equations governing the overall dynamics of the estimator without memory are given by
\begin{align}
& \dot{S}_t = d_f, \ \dot{\delta}_t = \delta_d, \ \dot{\tilde{x}} = \epsilon d_s, \nonumber \\
& \dot{\hat{W}}  = \epsilon F_{up,w}(\hat{W}, \hat{S}, e, \tilde{x}), \ \dot{\hat{S}}  = \epsilon F_{up,v}(\hat{W}, \hat{S}, e, \tilde{x}). 
\label{eq:estimator-dynamics-1}
\end{align}

Since memory as a function is useful only when the variations of the unknown function are faster than the update rate of the NN weights, the updates to $\hat{W}, \hat{S}, \hat{W}_m, \hat{S}_m$ are modeled to be slower relative to $S_t$. On the other hand, we model the key update to be faster because the memory can be effective only when the keys constantly reflect the changes. 
The equations governing the overall dynamics for the estimator with memory are as follows:
\begin{align}
&\dot{S}_t = d_f, \ \dot{\delta}_t = \delta_d, \ \dot{\tilde{x}} = \epsilon d_s, \nonumber \\
& \dot{\hat{W}}_m  = \epsilon F_{up,w}(\hat{W}_m, \hat{S}_m, e_m, \tilde{x}), \nonumber \\
& \dot{\hat{S}}_m  = \epsilon F_{up,v}(\hat{W}_m, \hat{S}_m, e_m, \tilde{x}),  \nonumber \\
& \dot{\mu} = F_{up,\mu}(\mu, k_\mu, \hat{W}_m, \hat{S}_m, e_m, \tilde{x}), \nonumber \\
& \dot{k}_{\mu} = F_{up, k}(k_\mu, \mu, \hat{W}_m, \hat{S}_m, e_m, \tilde{x}).
\label{eq:estimator-dynamics}
\end{align}

The set of all functions $(d_f, \delta_d)$ is denoted by $\mathcal{D}_f$. From now on we use $\sigma$, $\hat{\sigma}_m$, $\hat{\sigma}$ as shorthand notation for $\sigma(S_t^T\tilde{x})$, $\sigma(\hat{S}_m^T\tilde{x})$, $\sigma(\hat{S}^T\tilde{x})$. 

The timescale separation between the fast and slow dynamics permits the definition of a boundary layer dynamics and a reduced order system for the average dynamics. The response of the original system can then be analyzed by the response of the much simpler boundary layer and the average dynamics. 
The boundary layer dynamics for the respective estimators are defined below.
\begin{definition}
{\it (boundary layer dynamics)
\begin{align}
& \dot{\hat{W}}^{bl}_m = 0,  \dot{\hat{S}}^{bl}_m = 0, \dot{\tilde{x}}^{bl} = 0, \dot{S}^{bl}_t = d_f, \dot{\delta}^{bl}_t = \delta_d \nonumber \\
& \dot{\mu}^{bl} = F_{up, \mu}(\mu^{bl}, k^{bl}_\mu, \hat{W}^{bl}_m, \hat{S}^{bl}_m, e^{bl}_m, \tilde{x}^{bl}), \nonumber \\
& \dot{k}^{bl}_\mu = F_{up,k}(k^{bl}_\mu, \mu^{bl}, \hat{W}^{bl}_m, \hat{S}^{bl}_m, e^{bl}_m, \tilde{x}^{bl}) \nonumber \\
& (\text{estimator with memory}) \nonumber \\
& \dot{\hat{W}}^{bl} = 0,  \dot{\hat{S}}^{bl} = 0, \dot{\tilde{x}}^{bl} = 0  \nonumber\\
& \dot{S}^{bl}_t = d_f, \dot{\delta}^{bl}_t = \delta_d \ (\text{estimator without memory}) \nonumber
\end{align}}
\label{def:bdrylayerdyn}
\end{definition}
Here the superscript $bl$ is used to denote the boundary layer variables corresponding to the variables of the original system. The variables $\hat{W}^{bl}_m, \hat{S}^{bl}_m$ and $\tilde{x}^{bl}$ being the state variables corresponding to the slow dynamics, the time derivatives of these variables are set to zero in the boundary layer dynamics. 

Denote the Memory Read output corresponding to the boundary layer dynamics by $M^{bl}_r = M_r(\mu^{bl}, k^{bl}_\mu, \hat{S}^{bl}_m, \tilde{x}^{bl})$ and $\sigma^{bl} = \sigma((S^{bl}_t)^T\tilde{x}^{bl})$. We make the following assumption on the Memory Read output. 
\begin{assumption}
(Memory) {\it (i) $N \geq 2$; (ii) $\norm{M^{bl}_r - \sigma^{bl}} \leq \delta_m$; (iii) $M_r(\mu, k_\mu, \hat{S}_m, \tilde{x})$ is a weighted linear combination of the columns of $\mu$, where the weights add up to one and is a uniformly continuous function of $k_\mu, \hat{S}_m$ and $\tilde{x}$; (iv) for all $(d_f, \delta_d) \in \mathcal{D}_f$, $e_{m,a} \rightarrow 0 \Rightarrow M_r \rightarrow \hat{\sigma}_m$ or $M_r = \hat{\sigma}_m$ when $e_{m,a} = 0$}.
\label{ass:memory}
\end{assumption}
The final assumption on the Memory Read output specifies the steady state of the memory in terms of the Memory Read output. This simplifying assumption is needed to characterize the structure of the solution of the estimator with memory. Next, we present the main theorem. Please see the Appendix for Assumptions [3-6].
\begin{theorem}
{\it Suppose the two NN estimators given by Eq. \eqref{eq:est-without-memory} and Eq. \eqref{eq:update-NN-with-memory} are input-output equivalent to the $\overline{\delta}$-approximator of $f_t$ at $t = 0$. Suppose Assumptions 1, 3-6 hold and the $\overline{\delta}$-approximator is irreducible. Then there exists $\epsilon^{*} > 0$ such that for all $(d_f, \delta_d) \in \mathcal{D}_f$, $\delta_t$ and for all $\epsilon \in (0, \epsilon^{*}], \epsilon < 1$,
\begin{equation}
\vert e_m(t) \vert \leq \frac{\vert e^{bl,\text{max}} \vert}{1+\alpha} + O( \alpha(\overline{\delta}+\delta_m)) \ \forall t, \nonumber 
\end{equation}
where $e^{bl,\text{max}}$ is the maximum of $e^{bl}$ for the boundary layer dynamics of the estimator without memory.}
\label{thm:thm:est}
\end{theorem}

Please see the Appendix for the proof. The result clearly shows that if the error in the Memory Read output is small, i.e. $\delta_m \ll 1$, $\overline{\delta} = 0$ and if we set $\alpha = 1$, then the maximum deviation of the error for the estimator with memory is less than half of the maximum deviation of the error for the estimator without memory. The main purpose of the analysis is to provide theoretical ground and justification for the improvement in estimation from an external memory.

\section{Working Memory and Control Algorithm}
\label{sec:memint&contalg}


In this section, we provide a detailed description of the external working memory for the neural adaptive controller in Fig. \ref{fig:cwmem}. We then specify the control law and the NN update laws for the general case and the MRAC NN adaptive controller. Finally, we present a theorem that establishes the bounded stability of the proposed MRAC controller.

\subsection{External Working Memory}
\label{sec:mem-int}

The working memory we propose here is based on the ideas discussed in Section \ref{sec:est}. We use the same notation used in Section \ref{sec:est} for the state of the memory. The $i$-th column vector of the matrix $\mu$ is denoted by $\mu_i$. Let the number of columns be $n_s$. For an input vector $v$, the $i$-th element of the output of the softmax function, $\text{softmax}(v)_i = \frac{\exp(v_i)}{\sum_j \exp(v_j)}$. The estimated NN weight matrices are denoted by, $\hat{W} \in \mathbb{R}^{N\times m}$, $\hat{V} \in \mathbb{R}^{n\times N}$, $\hat{b}_v \in \mathbb{R}^{N}$, and $\hat{b}_w \in \mathbb{R}^{m}$, where $m$ is the dimension of the control input $u$ defined in Eq. \eqref{eq:claw}. The output of the estimated NN $u_{ad} \in \mathbb{R}^m$. Let the vector $q_\mu \in \mathbb{R}^{1\times m}$. 
For the proposed memory, the specific form of the Memory Write, Memory Read and the NN output are given by,
\begin{align}
 & \text{Memory Write:} \ \dot{\mu}_i = -z_i\mu_i +c_wz_ia +z_i\hat{W}q_\mu^T, \nonumber\\
 & \text{Memory Read:} \ M_r = \mu z, \ z = \text{softmax}(\mu^Tq), \nonumber\\
 & \text{NN Output:} \ u_{ad} = - \hat{W}^T\left(\sigma(\hat{V}^T\tilde{x} + \hat{b}_v) + M_r\right) - \hat{b}_w.
 \label{eq:memop}
\end{align}
where $q$ is the query, $a$ is the write vector, $z$ is the output of the addressing mechanism or the attention weights, $c_w$ is a constant, $q_\mu$ is a controller dependent term that is a function of the estimation error.\\

{\it Memory Write:} 
The right-hand side of Memory Write consists of three terms: (i) a forget term (the first term), (ii) an update term which is based on the new information from the write vector (the second term) and (iii) an additional update term (the third term). The first term in Memory Write equation allows the memory to forget its contents at the $i$th location at the rate $z_i$. This term is also critical for stability of the controller. The middle term in Memory Write equation updates the memory contents with the information from the write vector $a$. The third term allows the controller to quickly update the memory when the output error is large inspite of the update provided by the write vector. Thus, the third term plays a complementing role to the second term. 
We demonstrate the effect of the update by the third term in the simulations later. 

We set the write vector $a$ as the output of the hidden layer, i.e.,
\begin{equation}
a = \sigma(\hat{V}^T\tilde{x} + \hat{b}_v),
\label{eq:writevec}
\end{equation}
because this is the new information. The weights $\{z_i, \forall \ i\}$ determine which memory location is updated by the controller. In our design, the weights $z_i$s are set equal to a measure of similarity between the write vector $a$ (also the query as defined below) and the respective memory contents, which is then converted to a set of weights that add up to one through a softmax function \eqref{eq:memop}. 

{\it Memory Read:} 
We set the query vector to be the write vector itself, i.e.,
\beq q = \sigma(\hat{V}^T\tilde{x} + \hat{b}_v). \label{eq:query1} \eeq
The key for the respective memory locations are set to be the memory vectors themselves, i.e.,
\begin{equation}
k_i = \mu_i. \nonumber 
\end{equation}

The Memory Read output $M_r$ is a weighted combination of the memory vectors, where the weights for the respective locations are the same weights, $z_i$s, used for the write operation. 
The exact form of Memory Read $M_r$ is described in Eq. \eqref{eq:memop}. Clearly, with the weights given by $z_i$s, the output gives weight to those memory vectors whose key is similar to the current query. The design choice  for the query and key conditions that the Memory Read output is not very different from the estimate $\sigma(\hat{V}^T\tilde{x} + \hat{b}_v)$ of the actual hidden layer output.

{\it Modified NN Output:} the final output is computed by modifying the NN output by the information read from the memory as described in \eqref{eq:memop}. This modification provides an additional context to the learner that can potentially improve the compensation of the unknown part of the system dynamics as illustrated in the previous section.

\subsection{Control Algorithm}
\label{sec:controllaw}
First we present the complete set of equations for the general control architecture in Fig. \ref{fig:cwmem} and then present their specific form for the Model Reference Adaptive Control (MRAC) controller for a linear plant with matched uncertainty. 

\setcounter{subsubsection}{1}
\thesubsubsection \ {\it General Control and Update Law:} \label{sec:genmann}
The Memory Write equation, the Memory Read equation and the NN output are the same as the equations in \eqref{eq:memop}. The control input for NN adaptive control is a combination of a base controller $u_{bl}$, which is problem specific, the NN output $u_{ad}$ and a ``robustifying term'' $v$ \cite{lewis1996multilayer, lewis1998neural}. The final control input is given by,
\beq u = u_{bl} + u_{ad} + v, \nonumber \eeq

The variable $q_\mu$ in \eqref{eq:memop} is problem specific and depends on the Lyapunov function (without the NN error term). The NN update law, which constitutes the learning algorithm for the proposed architecture, is the regular update law for a two layer NN \cite{lewis1998neural},
\[
\left[\begin{array}{c} \dot{\hat{W}} \\ \dot{\hat{b}}^T_w \end{array}\right] = \gamma_w\left(\hat{\sigma} - \hat{\sigma}^{'}  \left(\hat{V}^T\tilde{x}+\hat{b}_v\right)\right)q_\mu - \kappa \gamma_w\norm{e}\left[\begin{array}{c} \hat{W} \\ \hat{b}^T_w \end{array}\right]  \nonumber
\]
\beq 
\left[\begin{array}{c} \dot{\hat{V}} \\ \dot{\hat{b}}^T_v \end{array}\right] = \gamma_v \left[\begin{array} {c} \tilde{x} \\ 1 \end{array} \right] q_\mu \left[ \begin{array}{c} \hat{W} \\ \hat{b}^T_w \end{array}\right]^T \hat{\sigma}^{'} - \kappa \gamma_v\norm{e}\left[\begin{array}{c} \hat{V} \\ \hat{b}^T_v \end{array}\right], 
\label{eq:NNupdate} \eeq
where $\gamma_v, \gamma_w$ and $\kappa$ are scalar gains and $e$ is the error.

\stepcounter{subsubsection}
\thesubsubsection \ {\it MRAC Controller:} 
\label{sec:mrac}
The system for a standard model reference adaptive control problem is given in \eqref{eq:sys-lmu}. 
\beq \dot{x} = Ax +B(u + f_t(x)) +B_r s \label{eq:sys-lmu} \eeq

The above equations represent the dynamics of the plant. This is a linear plant, whose system matrices $A$ and $B$ are known, and $f_t(x)$ is the matched nonlinear uncertainty. The signal $s$ is the command signal. 
The objective of the controller is to track the state of the reference model, given by $\dot{x}_{\text{ref}} = A_{\text{ref}} x_{\text{ref}} +B_r s$. The error in the tracking, $e = x - x_{\text{ref}}$. 


The control input and the update laws are the same as \eqref{eq:claw} and \eqref{eq:NNupdate}. The input to the NN is the state $x$ itself, i.e., $\tilde{x} = x$. The base controller input $u_{bl}$ is the standard Linear Quadratic Regulator (LQR) controller. The matrices $Q \in \mathbb{R}^{n\times n}$ and $R \in \mathbb{R}^{m\times m}$ that specify the LQR cost are given by, 
\begin{equation}
Q = K_v I_n, R = K_r I_m, \nonumber
\end{equation} 
where $I_l$ is the identity matrix of dimension $l$, $K_v, K_r$ are scalars and $K_v > 0, K_r > 0$. We denote the weights of the NN that is an approximation of $f(\cdot)$ with error $\epsilon$ by $W, V, b_w$ and $b_v$. From now on, we use a shorthand notation to represent a two layer NN with weights $W, V, b_w$ and $b_v$. The notation represents the NN by $W^T\sigma(V^Tx_e)$, where the biases $b_w$ and $b_v$ are implicit and included in $W$ and $V$ and $x_e = [\tilde{x}^T \ 1]^T$. We assume that the Frobenious norm of $Z = \text{diag}\{W, V\}$ for all possible $f$s is bounded by $Z_m$. The robustifying term in the control input is given by 
\begin{equation} 
v = - k_z\left(\norm{\hat{W}}_F + \norm{\hat{V}}_F + Z_{m}\right)\norm{e}_2, \label{eq:robterm}
\end{equation} 
where $k_z > 0$. The vector \[q_{\mu} = e^TPB,\] where $P$ is the matrix solution to the Lyapunov equation 
\begin{equation}
 A^T_{\text{ref}}P + PA_{\text{ref}} = -Q. \nonumber 
 \end{equation}

\subsection{Stability Theorem for MANN  Adaptive Controller}

In this section, we prove that the Memory Augmented Neural Network (MANN) MRAC controller leads to a closed loop system that is uniformly bounded. We first state the assumptions. 
\begin{assumption}
{\it We assume that (i) the reference signal $s$ is bounded and (ii) $\norm{Z_t}_F \leq Z_m$, where $Z_m$ is a constant, $Z_t =  \text{diag}\{W_t, V_t\}$ and $W_t$ and $V_t$ are the weights of the $\overline{\delta}$-approximator of $f_t$ within a compact subset of $\mathbb{R}^n$.}
\label{ass:boundedrefsig}
\end{assumption}
The following lemma establishes a bound on a certain term that appears in the stability analysis of the closed loop system. 
\begin{lemma}
{\it Let $\tilde{V}_t = V_t - \hat{V}$, $\tilde{W}_t = W_t - \hat{W}$, $Z_t = \text{diag}\{W_t, V_t\}$. Define, $w_1 = \tilde{W}_t^T\hat{\sigma}^{'}V_t^Tx_e + W_t^TO(\tilde{V}_t^Tx_e)^2  + \overline{\delta}$. Then, $\exists$ constants $b_1, b_2, b_3, c_2, c_3,$ such that
\[ \norm{w_1}_2 \leq b_1+ b_2\norm{\tilde{Z}_t}_F + b_3\norm{\tilde{Z}_t}_F\norm{e}_2, \nonumber \]
\[\norm{w_1}_2 \leq b_1+ b_2\norm{\tilde{Z}_t}_F + c_2Z_m\norm{e}_2 + c_{3}\norm{\hat{Z}_t}_F\norm{e}_2. \nonumber \]}
\label{lem:w1}
\end{lemma}
We present the stability result for the linear dynamical system  Eq. \eqref{eq:sys-lmu} and the MRAC controller in Section \ref{sec:mrac}. We consider the case where the $\overline{\delta}$-approximator (in Assumption \ref{ass:boundedrefsig}) of the unknown function $f_t$ of the system in Eq. \eqref{eq:sys-lmu} undergoes finite number of bounded abrupt changes. This is equivalent to the function $\tilde{f}$ defined by $\tilde{f}(t,\cdot) = f_t(\cdot)$ being discontinuous at instances when the function $f_t$ undergoes abrupt changes. Hence, we analyse the Caratheodory solution for the closed loop system.
\begin{definition}
\it Caratheodory solution of the closed loop system given by Eq. \eqref{eq:memop}, Eq. \eqref{eq:NNupdate}, Eq. \eqref{eq:claw} and Eq. \eqref{eq:sys-lmu} is the absolutely continuous function that satisfies the equations Eq. \eqref{eq:memop}, Eq. \eqref{eq:NNupdate}, and Eq. \eqref{eq:sys-lmu} for all $t$ except at the instant when $f_t$ undergoes abrupt changes.
\end{definition}

Next, we state the theorem for the boundedness of the Caratheodory solution of the closed loop system specified by the plant model \eqref{eq:sys-lmu} and the MRAC Controller defined in section \ref{sec:controllaw}.
\begin{theorem}
{\it Consider the plant model given by \eqref{eq:sys-lmu}, where the pair $(A,B)$ is stabilizable and the weights of the $\overline{\delta}$-approximator, $Z_t$, undergoes only abrupt changes. Let the controller be given by equations \eqref{eq:claw}, \eqref{eq:NNupdate} and \eqref{eq:memop}, where $u_{bl}$ is the stabilizing LQR controller of the system $(A,B)$ as described in Section \ref{sec:mrac}. Suppose\\
(i) Assumption \eqref{ass:boundedrefsig} is satisfied and the system state is available as feedback,\\
(ii) $\gamma_w = \gamma_v = K_v$, $\kappa = K_v^{0.75}$, $k_z \geq \text{max}\{c_2,c_3\}$, and gain $K_v$ is sufficiently large, and\\
(iii) $\norm{\Delta Z_t}_F \leq Z_m$, where $\Delta Z_t$ is the abrupt change in the $\overline{\delta}$-approximator and the number of abrupt changes are less than $\log{\gamma^{0.25}_v}$. \\
Then the Caratheodory solution of the closed loop system exists and is uniformly bounded.}
\label{thm:stability-1}
\end{theorem}
The analysis we present assumes or is restricted to the $\overline{\delta}$-approximator that maybe discontinuous at only finite number of points. A stability result for more general time varying functions (e.g., involving infinite number of discontinuities over a closed time interval) is much harder and is a subject for future work. In addition, stability result for the $\overline{\delta}$-approximator for more general class of systems besides the MRAC with the memory augmented NN is also a subject for future work.
%

%

\section{Discussion and Simulation Results}
\label{sec:sim-results}

In this section, we provide a detailed discussion and simulation results for the MRAC controller. In the simulation results that we present here, we provide 
(i) comparison with the response of the MANN controller without the first update term, i.e., when $c_w = 0$, 
(ii) comparison with the performance of a NN controller which has the same number of parameters as the MANN controller, where the number of parameters for the latter includes the size of the memory, (iii) comparison with the response of the MANN controller without the second update term 
and (iv) evidence for how memory augmentation improves learning.

\subsection{Flight Control Problem}

In this sub-section, we illustrate the memory augmented model reference adaptive controller in Fig. \ref{fig:cwmem} for the control of a flight's longitudinal dynamics. Denote the flight's angle of attack by $\alpha$, the flight's pitch by $q$ and the elevator control input by $u$. The flight's angle of attack and the pitch constitute the state of the system. The output of the system is its angle of attack, $\alpha$. In addition, we append an integrator, where the output of the integrator is the integral of the error between the output, i.e., the angle of attack and the command signal $s$ that the angle of attack has to track. Denote the output of the integrator by $e_I$, where $e_I = \int{\alpha - s}$. The system equations for the longitudinal dynamics appended with the output of the integrator is,
\[ \left[ \begin{array}{c} \dot{e}_I  \\ \dot{\alpha} \\ \dot{q} \end{array} \right] =  \left( \begin{array}{ccc} 0 & 1 & 0  \\ 0 & \frac{Z_\alpha}{mU} & 1 + \frac{Z_q}{mU}  \\ 0 & \frac{M_\alpha}{I_y} & \frac{M_q}{I_y} \end{array} \right)\left[ \begin{array}{c} e_I  \\ \alpha \\ q \end{array} \right]\nonumber \]
\beq+ \left( \begin{array} {c} 0 \\ \frac{Z_\delta}{mU} \\ \frac{M_\delta}{I_y} \end{array} \right)(u + f(x)) + \left( \begin{array} {c} -1 \\ 0 \\ 0 \end{array} \right)s. \eeq

The system parameters are that of Boeing-$747$ flight. We assume that the flight is traveling at a speed of $U = 274 \ \text{m/s}$ ($0.8$ Mach) and at an altitude of $h = 6000 \ \text{m}$. The flight's mass is $m = 288773 \ \text{Kg}$, and its moment of inertia $I_y = 44877574 \ \text{Kg}\text{m}^2$. The base controller is the LQR controller. The matrices that define the cost of the LQR controller are given by $Q = I$ and $R = 1$. The values for the other parameters in the system equation above are
\begin{align}
& \frac{Z_\alpha}{mU} = -0.32, 1 + \frac{Z_q}{mU} = 0.86, \frac{M_\alpha}{I_y} = -0.93, \nonumber \\
& \frac{M_q}{I_y}  = -0.43, \frac{Z_\delta}{mU} = -0.02,  \frac{M_\delta}{I_y} = -1.16 
\end{align}

In the results we provide here, the robustifying gain $k_z$ is set to $0$ to avoid the high frequency oscillations that occurs when it is set to a higher value. The learning rates or the gains in the NN update laws \eqref{eq:NNupdate} are set as $\gamma_w = \gamma_v = 10$ and $\kappa = 0$. The factor $c_w$ in the Memory Write equation is set as $3/4$. In the first and last two examples we consider here, the number of hidden layer neurons ($N$) of the respective neural networks are set as $4$ and then $5$ for the first and second example, respectively. The number of memory vectors in the external working memory, $n_s$, is set to $1$. 

We observed that the initial values of the NN weights influence the performance of the controller at least in the initial phase. Here we set the initial values of the weights in the outer layer of NN as $0$. The weights in the hidden layer of the NN and the elements of the memory vectors are randomly initialized to a number between $0$ and $1$. 
We now specify the examples we consider. In example $1$, the function $f(x)$ is given by:
\begin{align}
& f(x) = C_f(t)x^2, \ \text{where} \ C_f = 0.1 \ \text{at} \ t = 0,  \nonumber\\
& C_f \rightarrow 50 C_f \ \text{at} \ t = 5 \ \text{and} \ C_f \rightarrow 2 C_f \ \text{at} \ t = 25.
\end{align}

\noindent 
In example $2$, the function $f(x)$ is given by:
\begin{align}
& f(x) = x^2 + 0.1C_f, \ \text{where} \ C_f = 0.1 \ \text{at} \ t = 0,  \nonumber\\
& C_f \rightarrow 10 C_f \ \text{at} \ t = 5 \ \text{and} \ C_f \rightarrow 2 C_f \ \text{at} \ t = 25.
\end{align}

In example $3$, the function $f(x)$ goes through changes that are one order higher than that in example $2$:
\begin{align}
& f(x) = x^2 + 0.1C_f, \ \text{where} \ C_f = 0.1 \ \text{at} \ t = 0,  \nonumber\\
& C_f \rightarrow 100 C_f \ \text{at} \ t = 5 \ \text{and} \ C_f \rightarrow 2 C_f \ \text{at} \ t = 25
\end{align}

And in example $4$, the function $f(x)$ goes through changes that are not affine:
\begin{align}
& f(x) = x^2 + 0.1C_f, \ \text{where} \ C_f = 0.1 \ \text{at} \ t = 0,  \nonumber\\
& C_f \rightarrow \norm{x} \ \text{at} \ t = 5 \ \text{and} \ C_f \rightarrow 10\norm{x} \ \text{at} \ t = 25
\end{align}


\begin{figure}[htp]
\begin{tabular}{ll}
\includegraphics[scale = 0.2]{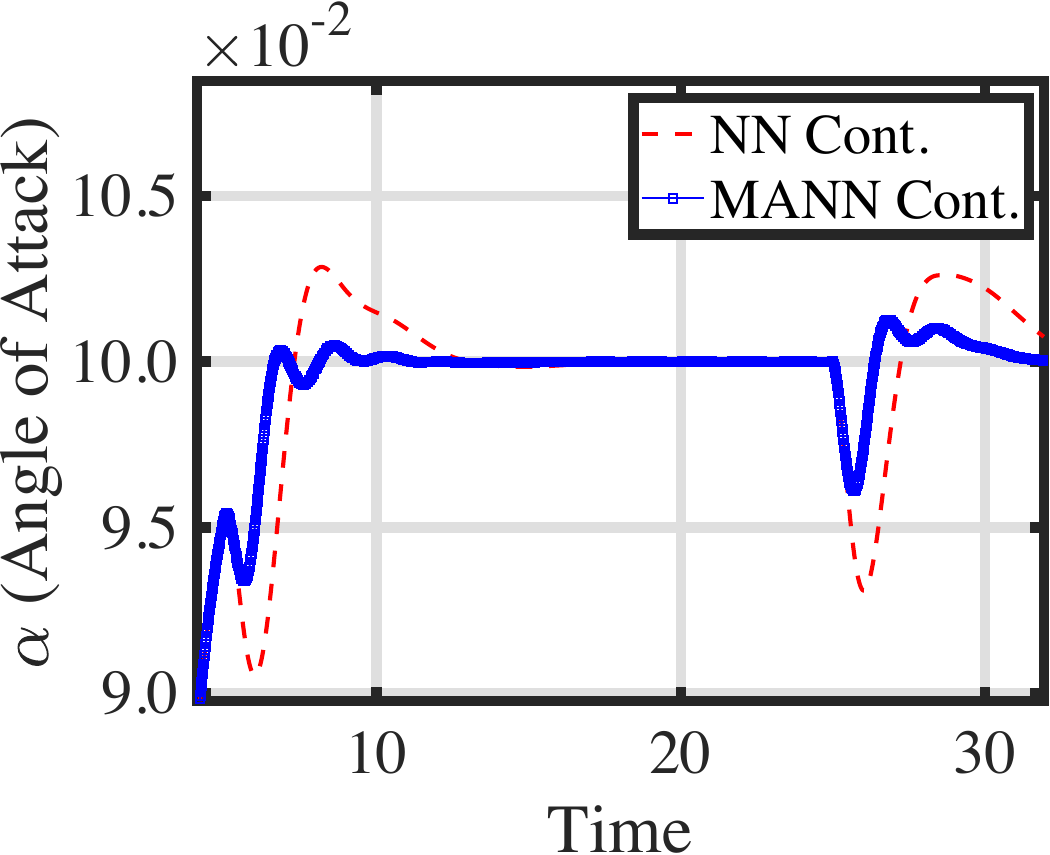} &\includegraphics[scale = 0.2]{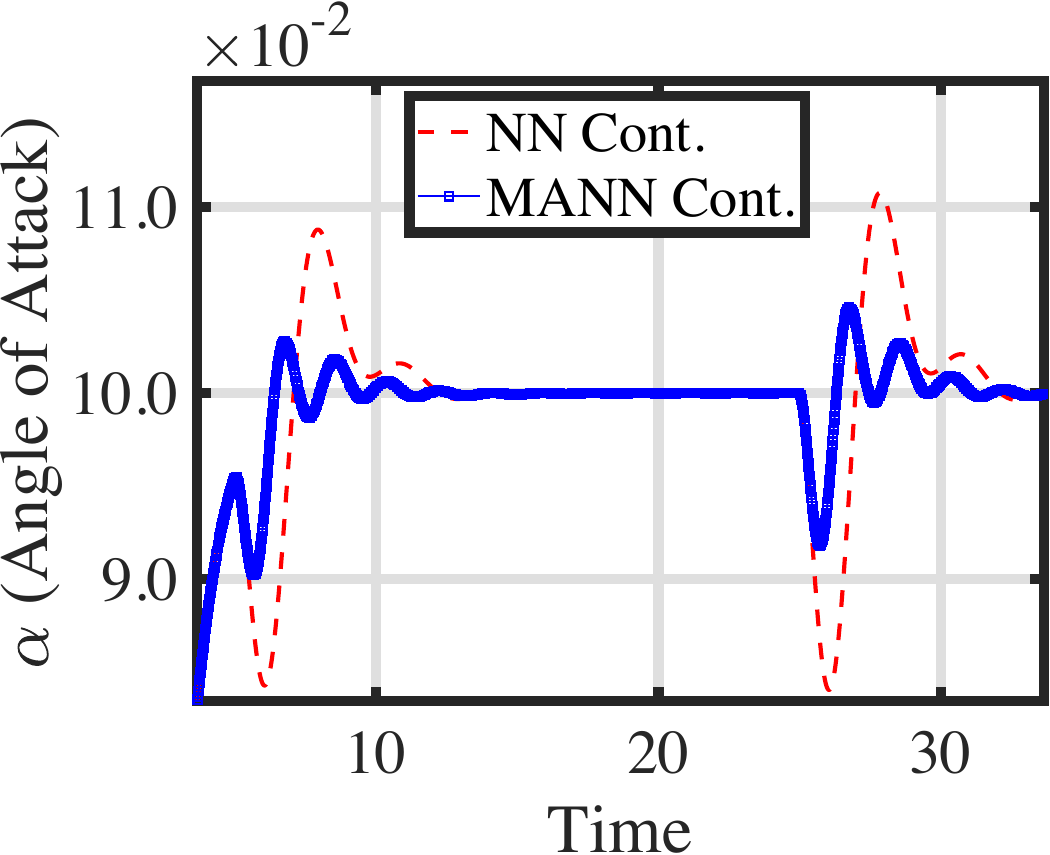} \\
\includegraphics[scale = 0.2]{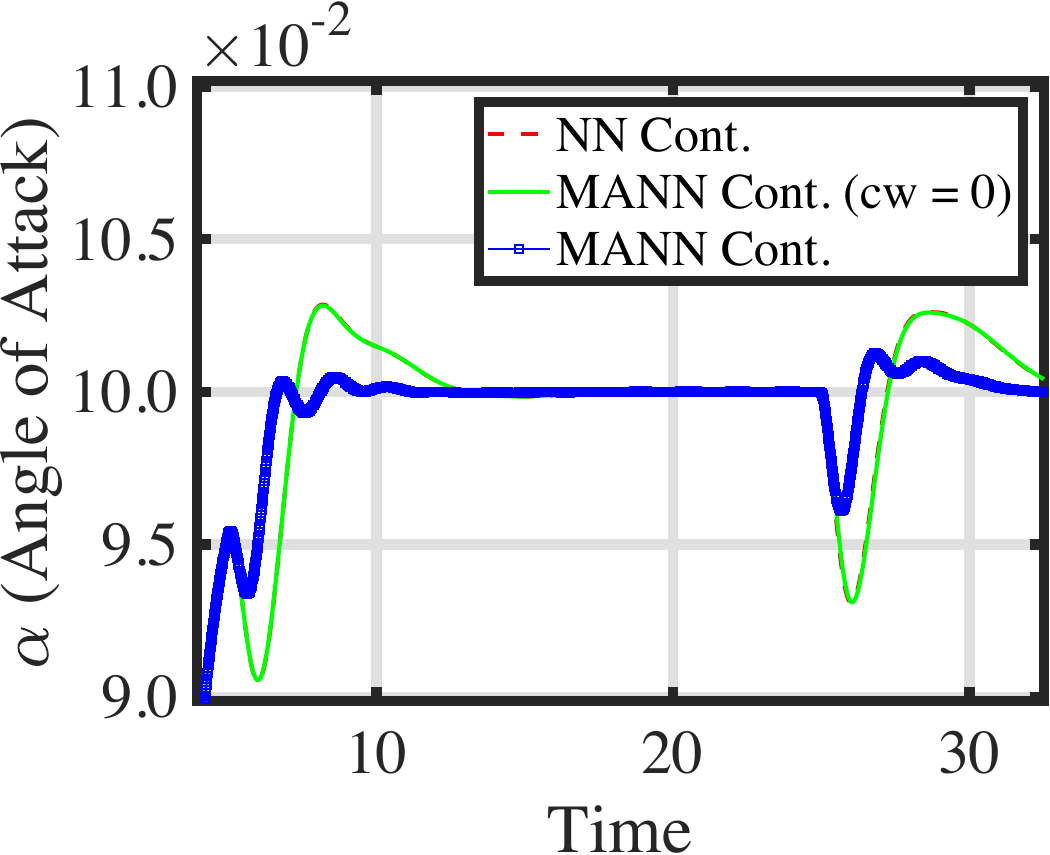}&\includegraphics[scale = 0.2]{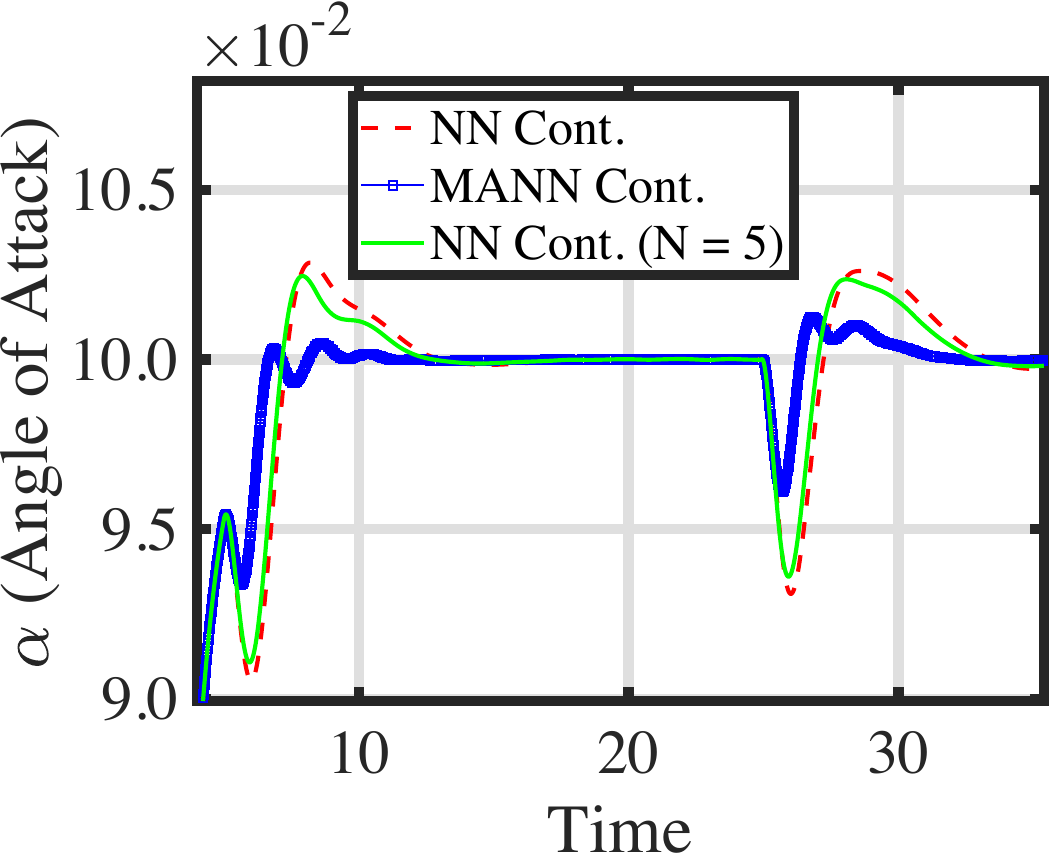} \\
\includegraphics[scale = 0.2]{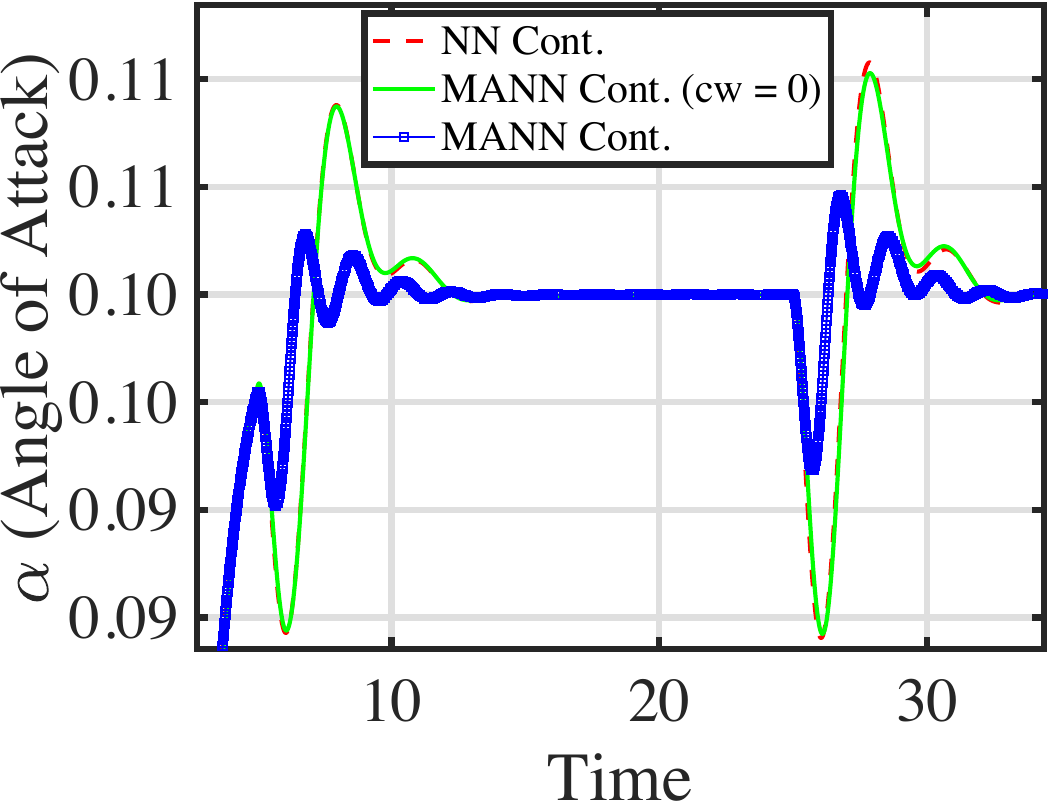}&\includegraphics[scale = 0.2]{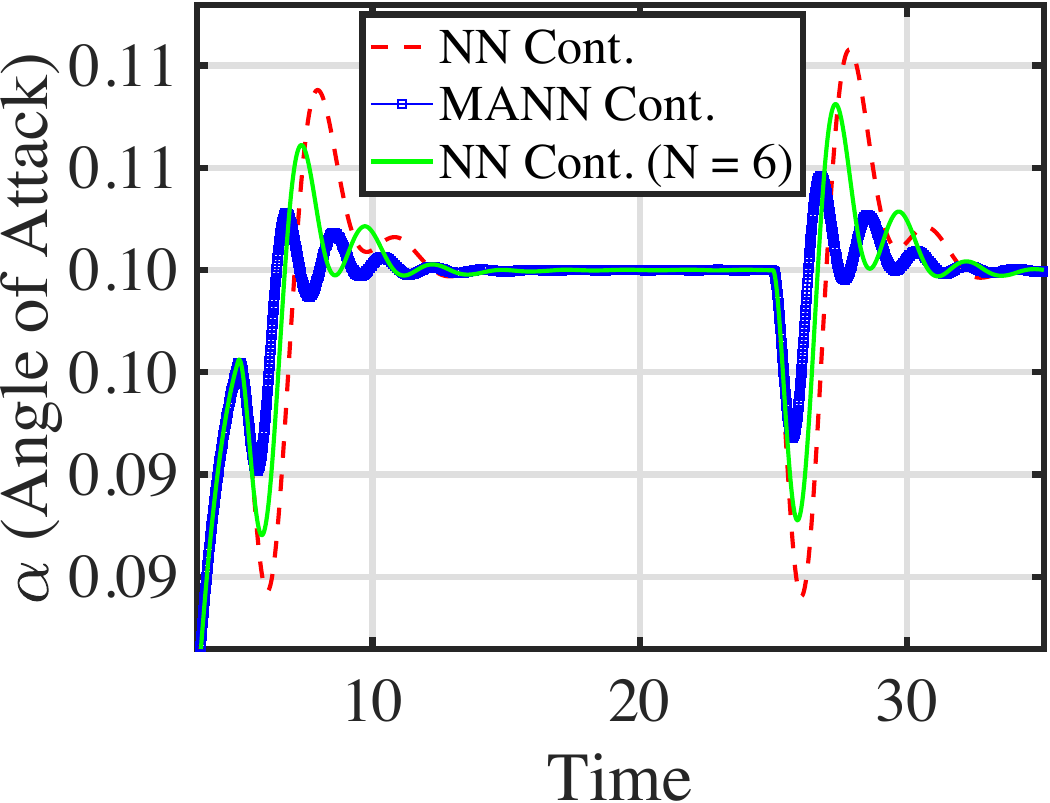}
\end{tabular}
\caption{Comparison of MRAC flight controllers with and without Memory. Left above: angle of attack response $\alpha$ (example 1), right above: angle of attack response $\alpha$ (example 2), left middle: comparison with MANN controller without the first update term (example 1), right middle: comparison with $N = 5$ NN controller (example 1), left below: comparison with MANN controller without the first update term (example 2), right below: comparison with $N = 6$ NN controller (example 2).}
\label{fig:flightcontrol}
\end{figure}

\begin{figure}[tp]
\begin{tabular}{ll}
\includegraphics[scale = 0.2]{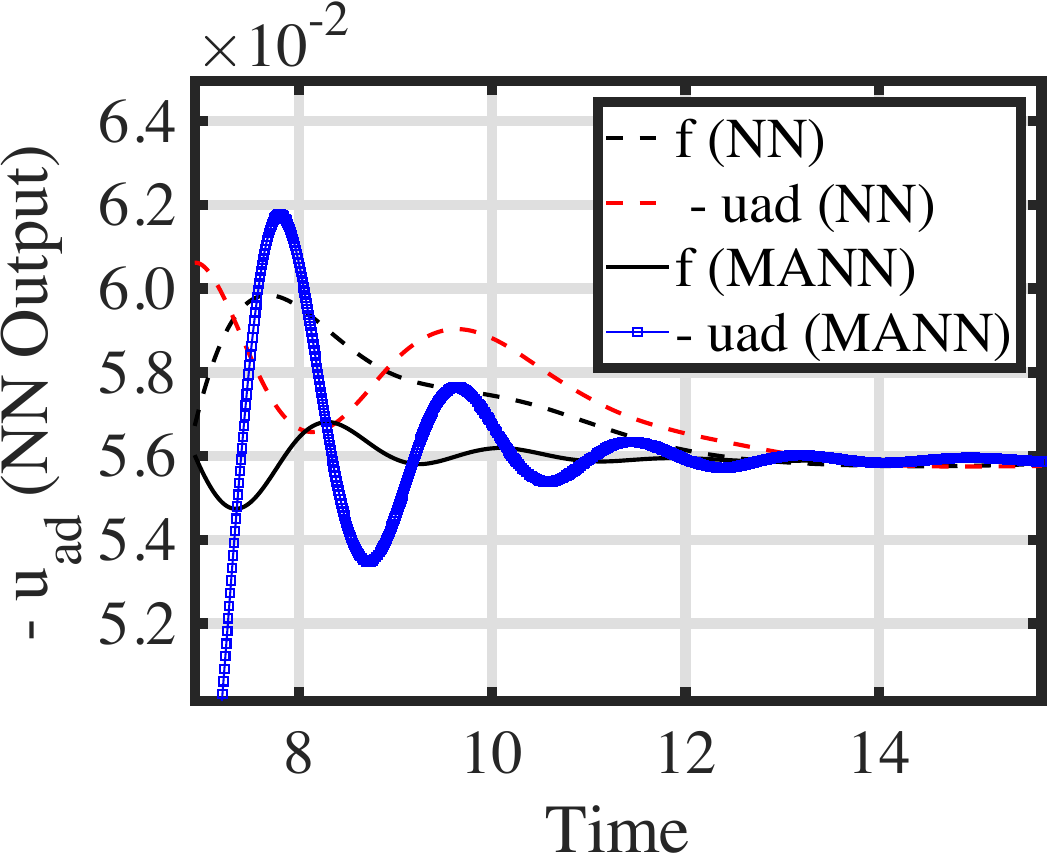}&\includegraphics[scale = 0.2]{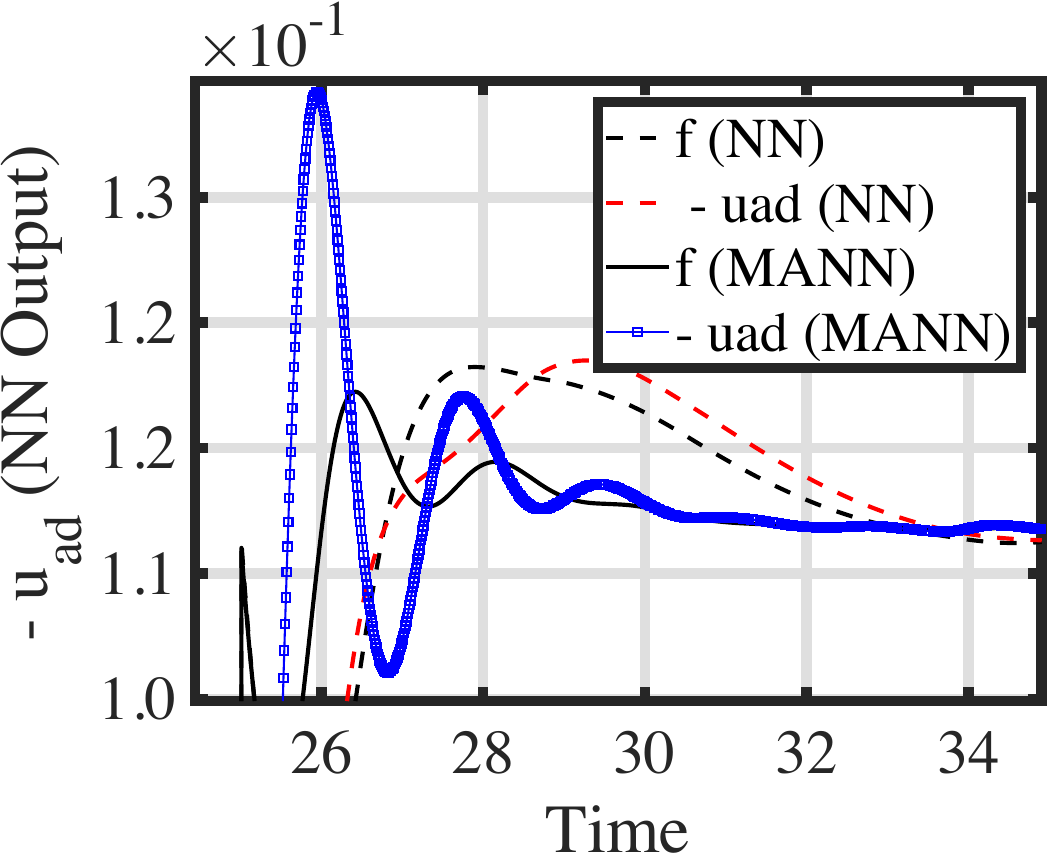} \\
\end{tabular}
\caption{Illustration of {\it induced learning} in the flight control problem. Left: Negative of NN output for both controllers at first abrupt change (example 1), right: Negative of NN output for both controllers at second abrupt change (example 1).}
\label{fig:flightcontrol-indlearning}
\end{figure}

\begin{figure}[htp]
\begin{tabular}{ll}
\includegraphics[scale = 0.2]{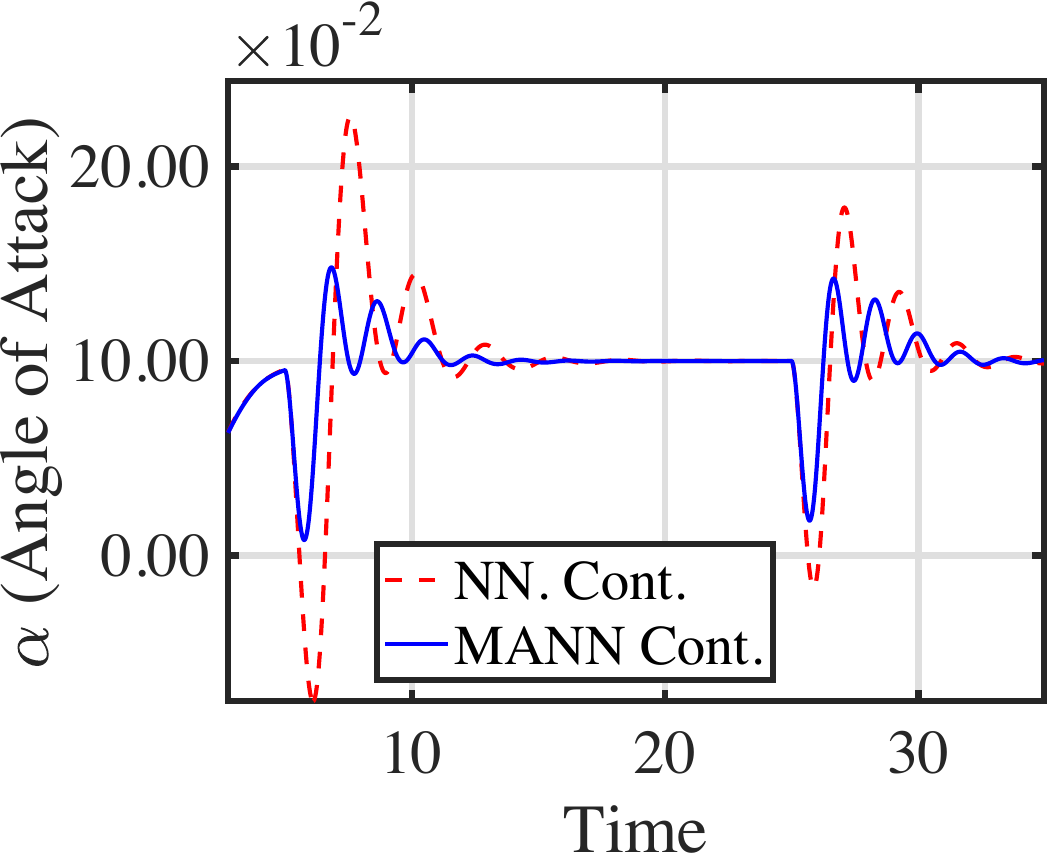} &\includegraphics[scale = 0.2]{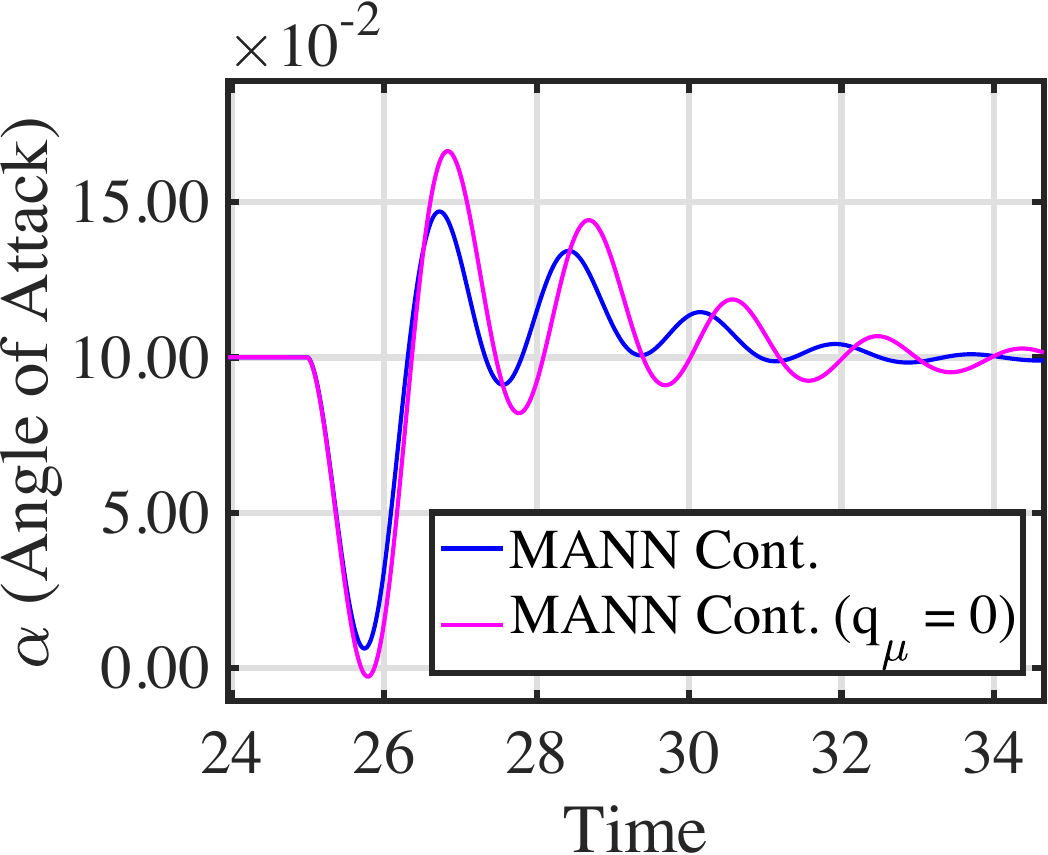} \\
\includegraphics[scale = 0.2]{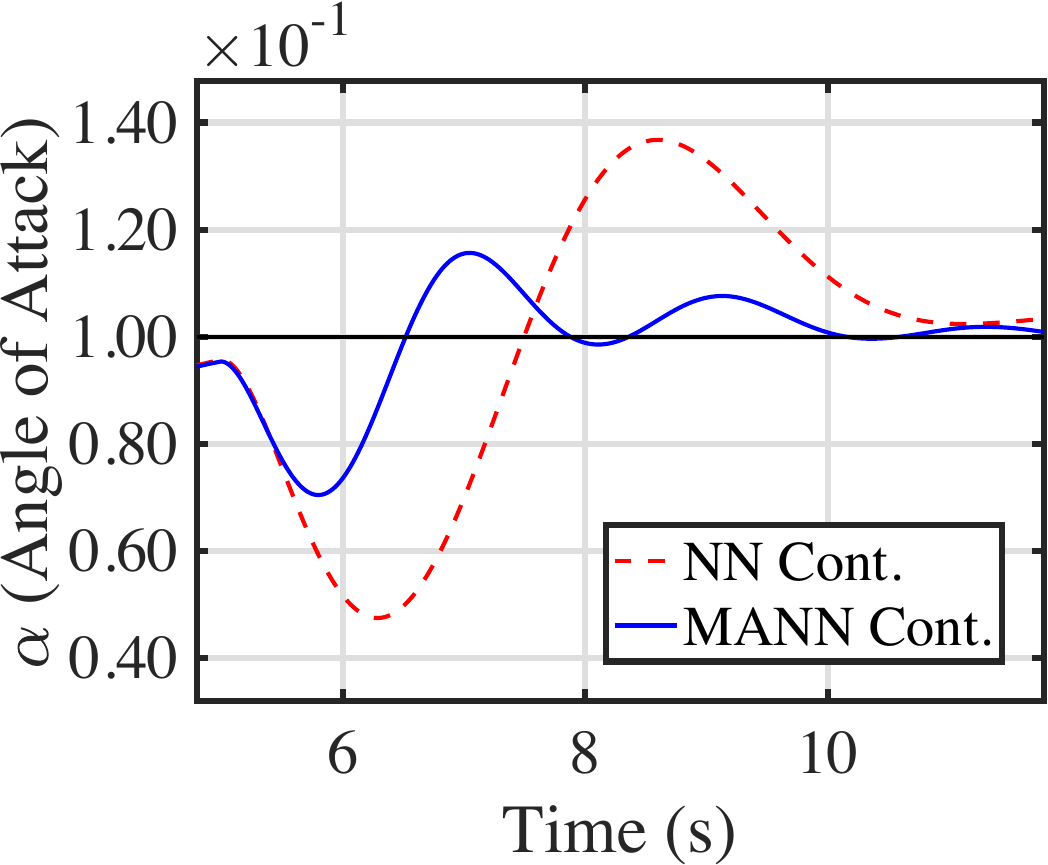} &\includegraphics[scale = 0.2]{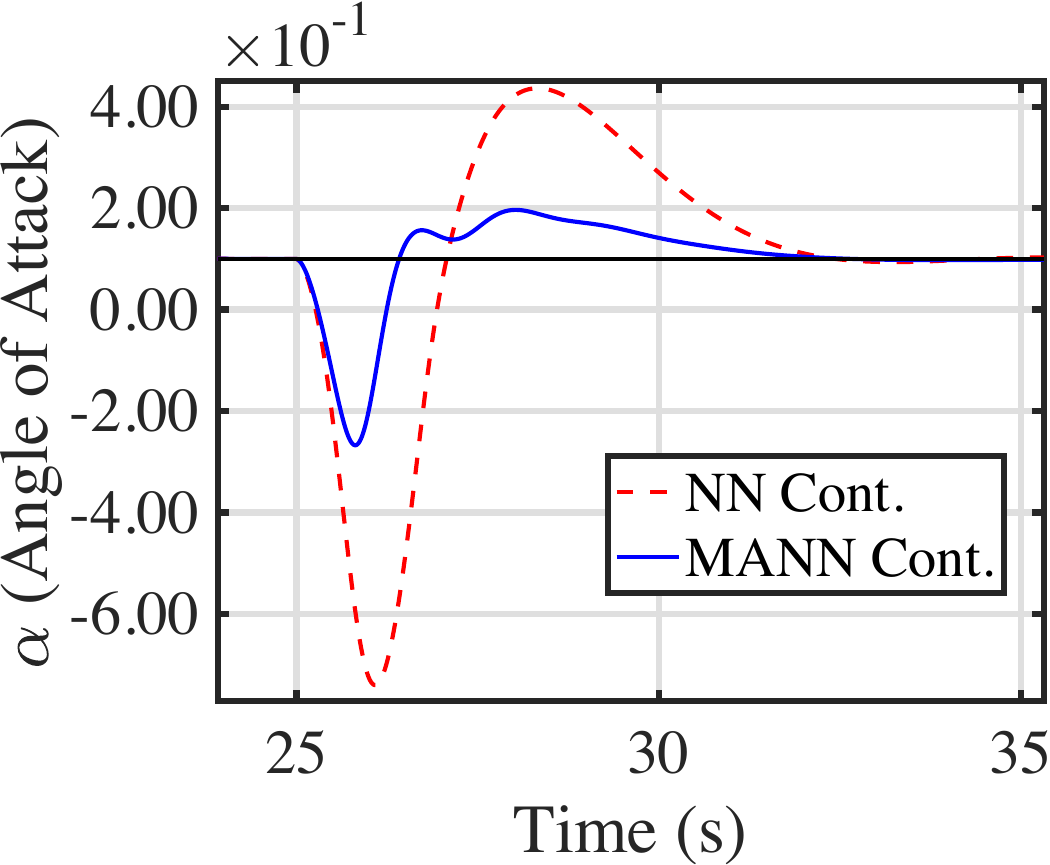} \\
\includegraphics[scale = 0.2]{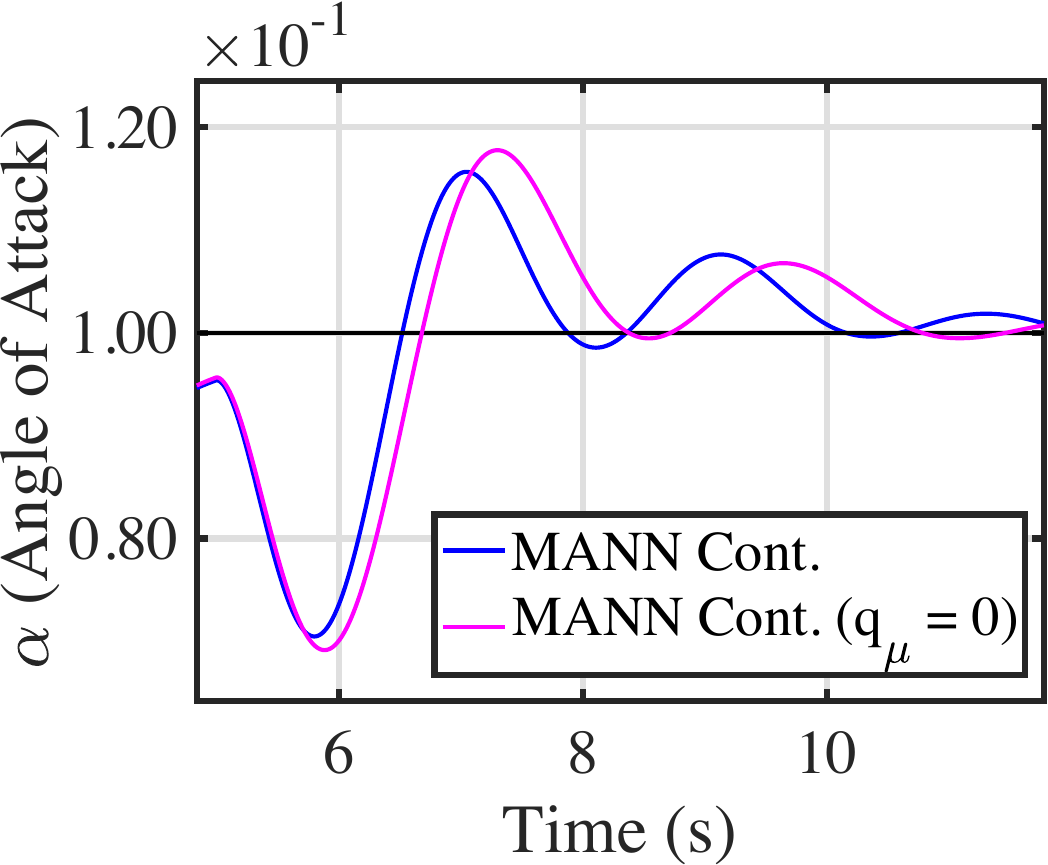} &\includegraphics[scale = 0.2]{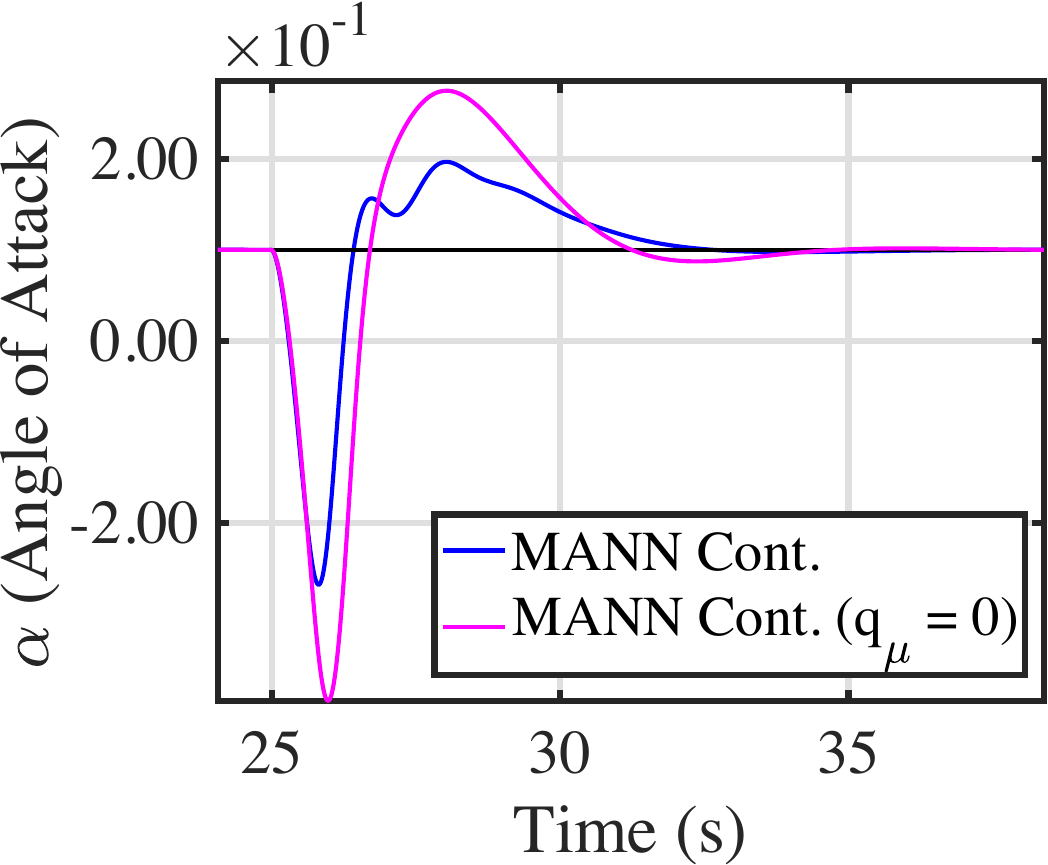} \\
\end{tabular}
\caption{Comparison of MRAC flight controllers with and without Memory. Left above: angle of attack response $\alpha$ (example 3), right above: comparison with MANN controller with $q_\mu = 0$ at the second abrupt change (example 3), middle left: angle of attack response $\alpha$ at first abrupt change (example 4), middle right: angle of attack response $\alpha$ at second abrupt change (example 4), bottom left: comparison with MANN controller with $q_\mu = 0$ at the first abrupt change (example 4), bottom right: comparison with MANN controller with $q_\mu = 0$ at the second abrupt change (example 4).}
\label{fig:flightcontrol-1}
\end{figure}

\begin{table}[h]
\centering
\caption{Flight Control, Peak Deviation, $ \max \vert \alpha - s \vert$}
\begin{tabular}{|c|c|c|}
\hline
Example & 1 & 2\\
\hline
NN cont. (I) &  $0.54^o$ ($N = 4$) & $0.89^{o}$ ($N = 5$)\\
\hline
NN cont. (II) &  $0.51^o$ ($N = 5$) & $0.7^{o}$ ($N = 6$) \\
\hline
MANN Cont. &  $0.38^o$  ($N = 4$) & $0.47^{o}$ ($N = 5$) \\
\hline
Reduction (from (II))&  25.5\% & 32.6\%\\
\hline
\end{tabular}
\label{table:peak}
\end{table} 

\begin{table}[h]
\centering
\caption{Flight Control, Settling Time (1 \% error) }
\begin{tabular}{|c|c|c|}
\hline
Example & 1 & 2  \\
\hline
NN cont. & $6.61$ s ($N = 4$) & $6.55$ s ($N = 5$)\\
\hline
NN cont. & $5.91$ s ($N = 5$) & $5.43$ s ($N = 6$)  \\
\hline
MANN Cont. & $3.45$ s ($N = 4$) & $4.1$ s ($N = 5$) \\
\hline
\end{tabular}
\label{table:settime}
\end{table} 

\begin{figure}[tp]
\begin{tabular}{ll}
\includegraphics[scale = 0.2]{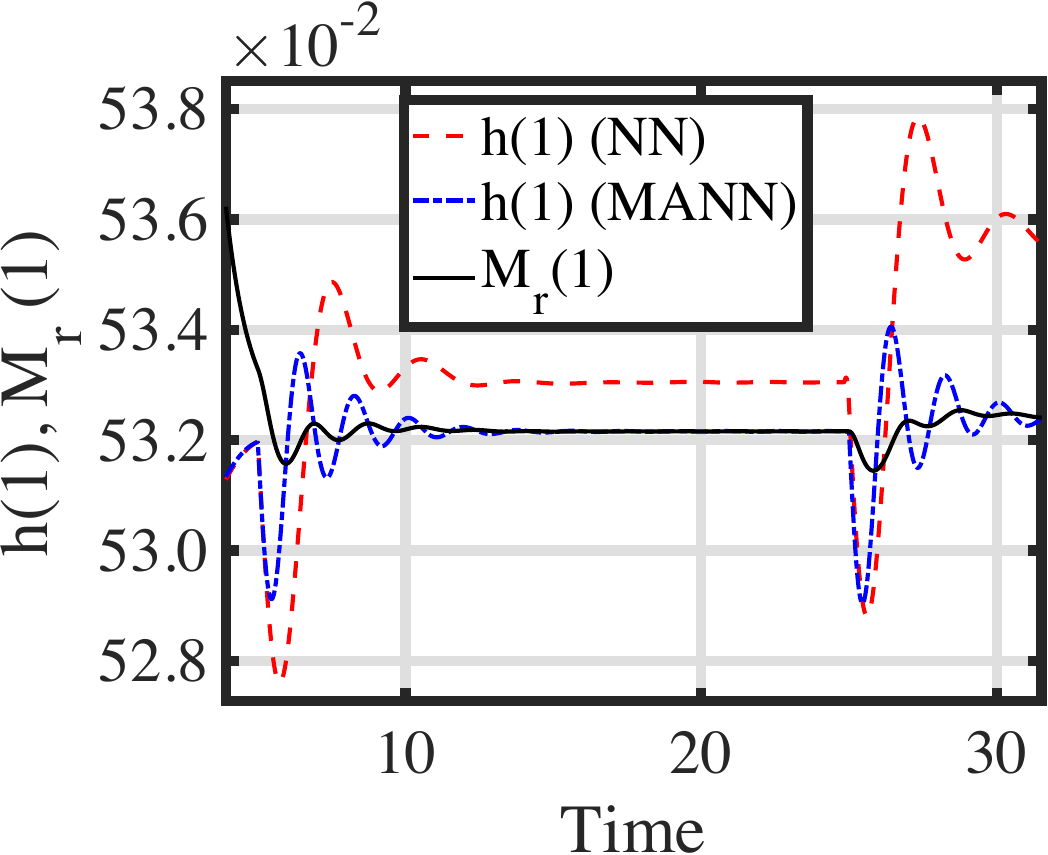} & \includegraphics[scale = 0.2]{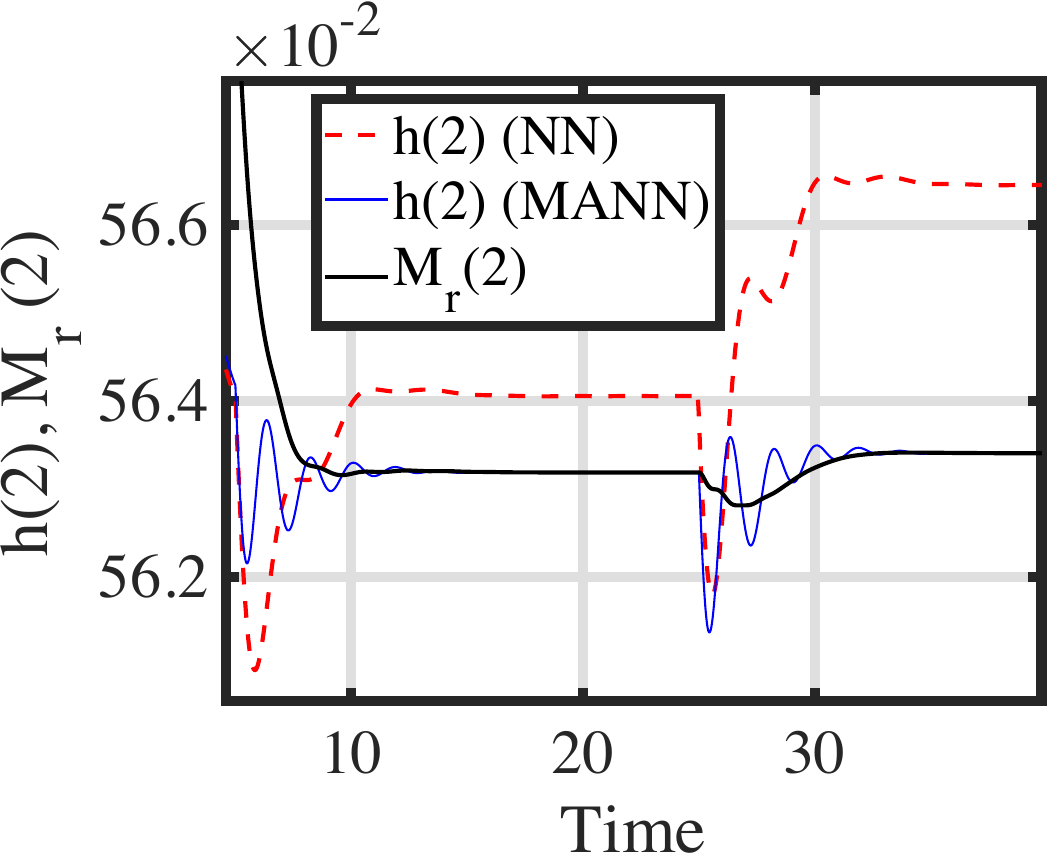}\\
\end{tabular}
\caption{Hidden layer output, $h$. Left: first component of hidden layer output (example 2), right: second component of hidden layer output (example 2).}
\label{fig:flightcontrol-repplot}
\end{figure}

\subsection{Discussion}

The top two plots of Fig. \ref{fig:flightcontrol} show the system output for the MANN controller and the NN controller without memory. We observe that the performance of the MANN controller is significantly better than the performance of the controller without memory both in terms of peak reduction and settling time. We emphasize that the examples or scenarios considered in these simulations capture diverse scenarios. Tables \ref{table:peak} and \ref{table:settime} provide values for two performance measures (i) peak deviation (ii) settling time for $1\%$ error. The peak deviation is the maximum deviation of the system output from the desired output and the settling time is the time for the error to settle within the stated error threshold. These metrics clearly reveal that the improvements obtained by the inclusion of a working memory are significant. The peak deviation is reduced significantly by $25\%$ and $32.6\%$ respectively in the two examples. We note that this is lower than the reduction predicted by Theorem \ref{thm:thm:est}, according to which, for $\alpha = 3/4$, the expected reduction is $\sim 42\%$. We hypothesize that the difference from the prediction of the theory arises from the error, $\delta_m$, of the Memory Read (see Section \ref{sec:est}).  

We also compare the performance of a NN controller that has the same number of parameters as the MANN controller, where the number of parameters in a MANN controller includes the number of memory components, which is $n_sN$. Through simple calculations we can show that this NN controller has $N = 5$ and $N = 6$ number of hidden layer neurons for examples $1$ and $2$ respectively. 
The response for this NN controller for the respective examples are shown in Fig. \ref{fig:flightcontrol}. 
Tables \ref{table:peak} and \ref{table:settime} provide the comparison of these controllers in terms of the performance measures. The measures clearly show that the MANN controller significantly outperforms the NN controller that has the same number of parameters. 

The bottom left and the middle left plots in Fig. \ref{fig:flightcontrol} provides the response for the MANN controller when its first update term is left out, i.e., when $c_w = 0$. We observe that the response without this term is not any better than the controller without memory. Figure \ref{fig:flightcontrol-indlearning} shows the plot of the function estimation error for example $1$. We observe that the function estimation error for the MANN controller reduces to zero faster when compared to the controller without memory. 

In Fig. \ref{fig:flightcontrol-repplot}, we provide evidence for how memory influences the learning. The plots show the first two components of the hidden layer output for the NN controller and the MANN controller and $M_r$ (the output of Memory Read \eqref{eq:memop}). In these plots, the Memory Read output $M_r$ is scaled by $1/c_w$ to account for the same factor in the first update term of the Write equation in \eqref{eq:memop}. 
From the plots we can conclude that the hidden layer output of the MANN controller converges nearer to the hidden layer value before the abrupt change, while, in contrast, it converges to a very different value for the controller without memory. This suggests that the memory is inducing the NN to converge to a network with very similar hidden layer weights. This combined with the observation that the function estimation error converges faster for the controller with memory (Figure \ref{fig:flightcontrol-indlearning}), is suggestive that the controller is being induced by the memory contents to find a good approximation in quick time. 

In Fig. \ref{fig:flightcontrol-1} we show the response of the MANN controller for example $3$ and example $4$. The response for example $4$ clearly shows that the MANN controller is able to respond faster than the regular NN controller even for non-affine changes. The top right plot and the bottom two plots compare the response of the MANN controller with that of the MANN controller without the second update term for the two examples. 
We observe that the MANN controller without the second abrupt term recovers slower than the MANN controller with the second update term for all the instances shown here. We note that the jumps in example $3$ are of a larger magnitude relative to the jumps in example $2$ and that the effect of the third update term is more pronounced in example $3$. This illustrates the importance of the second update term when the jumps are of a relatively larger magnitude. 

\section{Conclusion}
We proposed a novel control architecture for adaptive control of continuous time systems that is inspired from neuroscience. The proposed architecture augments an external working memory to the neural network that compensates the unknown nonlinear function in the system dynamics. We proposed a specific design for the working memory for this architecture. We formally argued that augmenting by an external working memory can improve the learning for a sub-problem of the adaptive control problem, an online estimation problem. Finally, we provided simulation results for a NN MRAC controller for linear systems with matched uncertainty. 
The simulations and the performance metrics clearly established that the controller augmented by an external working memory responds significantly faster and more accurately than its counterpart without the memory.

\bibliographystyle{IEEEtran}
\bibliography{Refs}

\section*{Appendix A: Proof of Lemma \ref{lem:app-capability-NNmodel}}

Since $f_{t} \in C(\mathcal{K})$, for any $\epsilon >  0$ and $x' \in \mathcal{K}$, there exists $\delta_1(\epsilon, t, x') > 0$ such that for any $x \in \mathcal{K}$, $\norm{x - x'} < \delta_1(\epsilon,t, x') \Rightarrow \norm{f_{t}(x) - f_{t}(x')} < \epsilon/2$. Similarly, since $F$ is continuous, for any $\epsilon >  0$ and $t' \in I$, there exists $\delta_2(\epsilon, t') > 0$ such that for any $t \in I$, $\vert t - t^{'} \vert < \delta_2(\epsilon,t') \Rightarrow \norm{f_t(x) - f_{t'}(x)} \leq \norm{f_t - f_{t'}}_{C(\mathcal{K})} < \epsilon/2$ for all $x \in \mathcal{K}$. 

For a given $(t',x') \in I \times \mathcal{K}$, let's set $\delta(\epsilon, t', x') = \min\{\delta_1(\epsilon,t', x'), \delta_2(\epsilon, t') \}$. Then, 
\begin{align}
& \norm{(t,x) - (t',x')} < \delta(\epsilon, t',x') \Rightarrow \vert t - t' \vert < \delta_2(\epsilon, t'), \nonumber \\
&  \norm{x - x'} < \delta_1(\epsilon, t', x'). \nonumber  
\end{align} 
Then, for a given $(t',x'), ~ \norm{(t,x) - (t',x')} < \delta(\epsilon, t',x')$ implies
\begin{align}
& \vert \tilde{f}(t,x) - \tilde{f}(t',x') \vert \leq \vert \tilde{f}(t,x) - \tilde{f}(t',x) \vert \nonumber \\
& + \vert \tilde{f}(t',x) - \tilde{f}(t',x') \vert \nonumber \\
& = \vert f_t(x) - f_{t'}(x) \vert + \vert f_{t'}(x) - f_{t'}(x') \vert \nonumber \\
&  < \epsilon/2 + \epsilon/2 = \epsilon. \nonumber 
\end{align}
Therefore, it follows that $\tilde{f}: I \times \mathcal{K} \rightarrow \mathbb{R}$ by $\tilde{f}(t,x) := f_t(x)$ is continuous. 

The fact that the nonlinear activation is the bounded sigmoid, it is continuous, bounded and non-constant. In addition, given that $I$ and $\mathcal{K}$ are compact, by Tychonoff's theorem $I \times \mathcal{K}$ is compact. Now Theorem 2 of \cite{hornik1991approximation} is applicable to $\tilde{f}$. Applying this theorem, it follows that, for any $\overline{\delta} > 0$, there exist $N \in \mathbb{N}$ and $w_i \in \mathbb{R}$, $\tilde{v}_i \in \mathbb{R}^{1\times (n+1)}, a_i \in \mathbb{R}$ such that 
\begin{equation}
\left\vert \tilde{f}(t,x) - \sum_{i=1}^N w_i \sigma(\tilde{v}_i [x;t] + a_i) \right\vert < \overline{\delta}. \nonumber 
\end{equation}
Let $\tilde{v}_i = [v_i, a^1_i]$, where $v_i$ is the first $n$ elements of $\tilde{v}_i$. Then, we can restate as
\begin{equation}
\left\vert \tilde{f}(t,x) - \sum_i w_i \sigma(v_i x + a_i + a^1_i t) \right\vert < \overline{\delta}. \nonumber 
\end{equation}
Let $W^T = [w_1, w_2,...., w_N], V^T = [v_1; v_2;....;v_N]$, and $b_t = [b^1_t, b^2_t, ...., b^N_t]$, where $b^i_t =  a_i + a^1_i t$. Then, we can further restate as 
\begin{equation}
\left\vert \tilde{f}(t,x) - W^T \sigma(V^T x + b_t) \right\vert < \overline{\delta}. \nonumber 
\end{equation}
That is, there exist constant $W \in \mathbb{R}^N$, $S^T_t = [V^T, b_t]$ such that  
\begin{equation}
\left\vert \tilde{f}(t,x) - W^T \sigma(S^T_t\tilde{x}) \right\vert < \overline{\delta}. \blacksquare \nonumber 
\end{equation}

\section*{Appendix B: Proof of Theorem \ref{thm:thm:est}}

First, we make the following observation as Lemma.
\begin{lemma}
{\it Suppose the two NN estimators given by Eq. \eqref{eq:est-without-memory} and Eq. \eqref{eq:update-NN-with-memory} are input-output equivalent to the $\overline{\delta}$-approximator of a function $f$, Assumption \ref{ass:memory} holds, then the set of all possible solutions for $\hat{W}_m, \hat{S}_m, \hat{W}, \hat{S}$ are given by
\begin{equation}
\hat{\sigma}_m = \sigma, \ \hat{W}_m = W/(1+\alpha), \ \hat{\sigma} = \sigma, \ \hat{W} = W \ \forall \ \tilde{x} \in \mathbb{R}^n \nonumber
\label{eq:est-sol} 
\end{equation}
or an identical permutation of the respective rows.}
\label{lem:est-mem-feas-sol}
\end{lemma}

{\it Proof}: Given that the estimator with memory is input-output equivalent to the $\overline{\delta}$-approximator
\beq e_{m,a} = W^T\sigma - \hat{W}^T_m(\hat{\sigma}_m + \alpha M_r) = 0 \ \forall \ \tilde{x}. \nonumber \eeq
Given that $\hat{\sigma}_m = M_r$ when $e_{m,a} = 0$ (Assumption \ref{ass:memory}), it follows that
\beq W^T\sigma - \hat{W}^T_m((1+\alpha)\hat{\sigma}_m) = 0  \ \forall \ \tilde{x}. \nonumber \eeq
Define $\tilde{W}_m$ by $\tilde{W}_m = (1+\alpha)\hat{W}$. Then the previous condition becomes
\beq W^T\sigma - \tilde{W}^T_m(\hat{\sigma}_m) = 0 \ \forall \ \tilde{x}. \label{eq:est-m-sscondn} \eeq 
This is essentially equivalent to a regular estimator with weights $\tilde{W}^T_m$ and hidden layer $\hat{\sigma}_m$. From the equivalence result for NNs with a sigmoid function as an activation function (Theorem 3.8 from \cite{kuurkova1994functionally}), and the fact that the $\overline{\delta}$-approximator is irreducible, if two networks are functionally equivalent as in \eqref{eq:est-m-sscondn}, then the weight matrices of the two NNs are either identical or the rows of matrices of one is a permutation of the other. Thus, the solution is $\hat{W}_m = W/(1+\alpha)$ and $\hat{\sigma}_m =  \sigma$ or an identical permutation of the respective rows. Similarly, for the estimator without memory the solution is $\hat{W} = W$ and $\hat{\sigma} =  \sigma$ or an identical permutation of the respective rows. $\blacksquare$

We first introduce a set of assumptions for the dynamical system described by Eq. \eqref{eq:estimator-dynamics}. Let 
\begin{align}
& x^T_s = [\hat{W}^T_m, \text{vec}(\hat{S}_m)^T, \tilde{x}^T], \nonumber \\
& x^T_f = [\text{vec}(\mu)^T, \text{vec}(k_\mu)^T, \text{vec}(S_t)^T, \delta_t], \ z^T = [x^T_s, x^T_f], \nonumber \\
& z^T_{bl} = [(\hat{W}^{bl}_m)^T, \text{vec}(\hat{S}^{bl}_m)^T, (\tilde{x}^{bl})^T, ...\nonumber \\
& \text{vec}(\mu^{bl})^T, \text{vec}(k^{bl}_\mu)^T, \text{vec}(S^{bl}_t)^T, \delta^{bl}_t], \nonumber
\end{align}
where $\text{vec}(\cdot)$ denotes the vectorial expansion of the input matrix, where the first sub-vector is the first column, second sub-vector is the second column and so on. We note that the state $x_s$ denotes the state of the slow dynamics. Let $\mathcal{R}_s$ denote the set of all possible values that $x_s$ can reach, $x_s(0) \in \mathcal{K}_s$ and $z(0) \in \mathcal{K}_f$. Let $\mathcal{S}_{bl}(z_{bl}(0), \tilde{d}_f)$ denote the solution space of the boundary layer dynamics (estimator with memory) starting from $z_{bl}(0)$ with the disturbance function given by $d_f$. Let $\omega_{f,o}(z) = \norm{y - y_m}$. Let $\omega_{f,i}$ and $\omega_{av,i}$ be two input measuring functions given by $\omega_{f,i}: \mathcal{V}_f \rightarrow \mathbb{R}_{\geq 0} \cup \{\infty\}$, $\omega_{av, i}: \mathcal{V}_f \rightarrow \mathbb{R}_{\geq 0} \cup \{\infty\}$. Measuring functions are functions that take values in $\mathbb{R}_{\geq 0} \cup \{\infty\}$ and need not be continuous. Next we define the average dynamics for the estimator dynamics given by Eq. \eqref{eq:estimator-dynamics}.
\begin{definition}
{\it (Admissible Average) The tuple $(m_g, \omega_{av,i}, \omega_{f,i}, F_{av}, \mathcal{R}_s, \mathcal{K}_f)$ constitutes an admissible average of system governed by Eq. \eqref{eq:estimator-dynamics}, if for each $\rho > 0$, there exist $T^{*} > 0$, and $\epsilon^{*} > 0$ such that for each 
\begin{align}
& T \geq T^{*}, \ \epsilon \in \left(0, \epsilon^{*}\right.], \ x_s \in \mathcal{R}_s, \ z^T_{bl} = [x^T_s, x^T_f] \in \mathcal{K}_f, \nonumber \\
& \tilde{d}_f \in \mathcal{D}_f, \omega_{f,o}(z_{bl}) \leq \norm{\omega_{f,i}(\tilde{d}_f)}_{\infty}, \ \phi_{bl} \in \mathcal{S}_{bl}(z_{bl}, \tilde{d}_f), \nonumber 
\end{align}
$\exists$ a measurable function $g: \mathbb{R}_{\geq 0} \rightarrow \mathbb{R}^{m_g}$, $\norm{g}_\infty \leq \norm{\omega_{av,i}(\tilde{d}_f)}_\infty$ such that
\begin{align}
& \left\vert \int_{0}^{T} F_\delta dt \right\vert \leq T\rho\epsilon, \  F_\delta = F_s(\phi_{bl}(t), d_s(t), \tilde{d}_f(t), \epsilon) \nonumber \\
& - \epsilon F_{av}(x_s, d_s(t), \tilde{d}_f(t), g(t)), \nonumber
\end{align}
where $F^T_s(x_s, x_f, d_s, \tilde{d}_f, \epsilon) = [\epsilon F^T_{up,w}(.), \epsilon F^T_{up,v}(.), \dot{\tilde{x}}(.)^T]$. The corresponding average system dynamics is given by 
\begin{equation}
\dot{x}^{av}_s = \epsilon F_{av}(x^{av}_s, d_s, \tilde{d}_f, g) \nonumber 
\end{equation}
where $x^{av}_s$ is the state of the average system dynamics.}
\label{def:time-av-F}
\end{definition}
Let $F^T_f(x_s, x_f, d_s, \tilde{d}_f, \epsilon) = [F^T_{up,\mu}(\cdot), F^T_{up,k}(\cdot), \text{vec}(\dot{S}_t)^T, \dot{\delta}_t]$. We make the following assumption on the disturbances $d_s$ and $d_f$. 
\begin{assumption} {\it (Disturbances)
(i) the set of all possible $d_s$ and $\tilde{d}_f$ are invariant to shift in time (ii) for each $T > 0$ and $\rho > 0$ there exists $\epsilon^{*} > 0$ such that for all $\epsilon \in (0, \epsilon^{*}]$ and for all $d_s$, $\norm{\epsilon(d_s(t) - d_s(0))} \leq \rho \ \forall \ t \in [0,T]$. }
\label{ass:disturbance}
\end{assumption}

The following assumption is on the boundary layer dynamics and the average dynamics.
\begin{assumption} {\it (Boundary Layer and Average Dynamics Stability)
(i) there exist a class $\mathcal{K}\mathcal{L}$-function (a continuous function that is zero at zero and strictly increasing in its first argument and decreasing to zero in its second argument) $\beta_f$ such that for all $z_{bl}(0) \in \mathcal{K}_f$ and all disturbances $\tilde{d}_f \in \mathcal{D}_f$, the solutions of boundary layer dynamics (Definition \ref{def:bdrylayerdyn}) exist and satisfies $\omega_{f,o}(z_{bl}(t)) \leq \max\left\{\beta_f(\omega_{f,o}(z_{bl}(0)),t), \norm{\omega_{f,i}(\tilde{d}_f)}_\infty\right\}$ (ii) admissible average system exists (iii) there exist a class $\mathcal{K}\mathcal{L}$-function $\beta_s$, an uniformly continuous measuring function $\omega_{s,o}: \mathcal{R}_s \rightarrow \mathbb{R}_{\geq 0} \cup \{\infty\}$, and a measuring function $\omega_{s,i}: \mathcal{V}_s \times \mathcal{V}_f \rightarrow \mathbb{R}_{\geq 0} \cup \{\infty\}$ such that for all $\epsilon > 0$, all $x_s(0) \in \mathcal{K}_s$, all disturbances $d_s$, $\tilde{d}_f \in \mathcal{D}_f$, all $\norm{g}_\infty \leq \norm{\omega_{av,i}(\tilde{d}_f)}_\infty$ the solutions of average system exist and satisfies $\omega_{s,o}(x^{av}_s(t)) \leq \text{max}\left\{\beta_s(\omega_{s,o}(x^{av}_s(0)),\epsilon t), \norm{\omega_{s,i}(d_s, \tilde{d}_f)}_\infty\right\}$.}
\label{ass:bdrylayer-average-dynamics}
\end{assumption} 
 
The following assumption is on the sets $\mathcal{K}_s$ and $\mathcal{K}_f$.
\begin{assumption} {\it (Initial Set)
(i) $x_s(0) \in \mathcal{K}_s, z(0) \in \mathcal{K}_f$, (ii) $\sup_{z \in \mathcal{K}_f} \omega_{f,o}(z) =: c_{f,o} < \infty$, $\sup_{x_s \in \mathcal{K}_s} \omega_{s,o}(x_s) =: c_{s,o} < \infty$, $\sup_{\tilde{d}_f \in \mathcal{V}_f} \omega_{f,i}(\tilde{d}_f) =: c_{f,i} < \infty$, $\sup_{d_s \in \mathcal{V}_s, \tilde{d}_f \in \mathcal{V}_f} \omega_{s,i}(d_s, \tilde{d}_f) =: c_{s,i} < \infty$ (ii) There exists $\alpha > 0$ such that a) $\{x_s : \omega_{s,o}(x_s) \leq c_{s,i} + \alpha,\} \subseteq \mathcal{K}_s$, b) $\{z \vert \omega_{f,o}(z) \leq c_{f,i} + \alpha, \omega_{s,o}(x_s) \leq \max\{\beta_s(\omega_{s,o}(x_s(0)),0), c_{s,i}\} + \alpha \} \subseteq \mathcal{K}_f$. }
\label{ass:initialset}
\end{assumption} 
  
The following assumption is on the continuity of the functions $F_{s}, F_f, F_{av}$ as defined above.
\begin{assumption}  {\it (Continuity)
Given $\mathcal{X}_s = \{x_s: \omega_{s,o}(x_s) \leq \max\{\beta_s(c_{s,o},0), c_{s,i}\}\}$, there exists $L > 0, M > 0, \eta > 0$ such that (a) $\mathcal{X}_s + \eta \mathcal{B} \subseteq \mathcal{R}_s$ (b) $\omega_{f,o}$ is uniformly continuous on $\mathcal{U}_f(\eta) := \{z: x_s \in  \mathcal{X}_s + \eta \mathcal{B}, x_f \in \mathcal{Z}_f + \eta \mathcal{B}\}$, $\mathcal{Z}_f = \{z: \omega_{f,o}(z) \leq \max\{\beta_f(c_{f,o},0), c_{f,i}\}\}$ (c) for each $\rho > 0$ there exists $\epsilon^{*} > 0$ such that for all $\tilde{d}_f \in \mathcal{D}_f, \norm{g}_\infty \leq \norm{\omega_{av,i}(\tilde{d}_f)}_\infty, [x^T_s, x^T_f]^T \in \mathcal{U}_f(\eta), [y^T_s, y^T_f]^T \in \mathcal{U}_f(\eta), \norm{d_s - w_s} \leq \epsilon^{*}, \epsilon \in (0,\epsilon^{*}]$,
\begin{align}
& \norm{F_s(x_s, x_f, d_s, \tilde{d}_f, \epsilon)} < \epsilon M, \norm{F_{av}(x_s, d_s, \tilde{d}_f, g)} < M, \nonumber \\
& \max\{\norm{x_s - y_s},\norm{x_f - y_f}\} \leq \epsilon^{*} \nonumber \\
& \Rightarrow \norm{F_s(x_s, x_f, d_s, \tilde{d}_f, \epsilon) - F_{s}(y_s, y_f, w_s, \tilde{d}_f, \epsilon)} \leq \epsilon \rho, \nonumber \\
& \norm{x_s - y_s} \leq \epsilon^{*} \Rightarrow\nonumber \\
&  \norm{F_f(x_s, x_f, d_s, \tilde{d}_f, \epsilon) - F_{f}(y_s, y_f, w_s, \tilde{d}_f, \epsilon)} \nonumber \\
& \leq L \norm{x_f - y_f} + \rho, \nonumber \\
&  \norm{F_{av}(x_s, d_s, \tilde{d}_f, g) - F_{av}(y_s, w_s, \tilde{d}_f, g)} \nonumber \\
& \leq L \norm{x_s - y_s} + \rho. \nonumber
\end{align}  }
\label{ass:continuity}
\end{assumption}

{\it Proof of the main theorem}: As a first step we show that: for each $\rho > 0$ there exists $\epsilon^{*} > 0$ such that for each $c \in [0, c_{f,i}]$, if $\omega_{f,o}(z) \leq c + \epsilon^{*}$ then there exists $z_c$ such that $\omega_{f,o}(z_c) \leq c$ and $\norm{z - z_c} \leq \rho$ (s1). For the estimator with memory $\omega_{f,o}(z) = \norm{e_m}$, where $e_m = y - y_m$.

For a given $c \in [0, c_{f,i}]$, if $\omega_{f,o}(z) \leq c$, then picking $z_c = z$ trivially proves statement s1 for this case. If $\omega_{f,o}(z_c) > c$, let $z^T = [\hat{W}^T_m, \text{vec}(\hat{S}_m)^T, \tilde{x}^T, \text{vec}(\mu)^T, \text{vec}(k_\mu)^T, \text{vec}(S_t)^T]$, and $z_c = [\hat{W}^T_{m,c}, \text{vec}(\hat{S}_m)^T, \tilde{x}^T, \text{vec}(\mu_c)^T, \text{vec}(k_\mu)^T, \text{vec}(S_t)^T]$. Note that the only difference between $z$ and $z_c$ are the variables $\hat{W}^T_m$ and $\text{vec}(\mu)$. Let $\Delta\hat{W}_{m,c} = \hat{W}_{m,c} - \hat{W}_m$ and $\Delta \mu = \mu_c - \mu$ with all columns being equal to $\Delta \mu_v$. Denote the unit vector along a vector $v$ by $Un(v)$. Denote the set of unit vectors perpendicular to a vector $v$ by $\mathcal{S}^{\perp}(v)$. Because $N \geq 2$, $\vert\mathcal{S}^{\perp}(v)\vert \geq 2$. Let $\tilde{\sigma}_m = \hat{\sigma}_m+M_r$. For a given $0 \leq \tilde{\rho} \leq \rho$, let 
\begin{align}
& \Delta \mu_v = \left\{ \begin{array}{cc} \in \frac{\tilde{\rho}}{\sqrt{2n_s}}\mathcal{S}^{\perp}(\tilde{\sigma}_m), \hat{W}^T_m\Delta \mu_v \geq 0 & \text{if} \ y > y_m \\ \in \frac{\tilde{\rho}}{\sqrt{2n_s}}\mathcal{S}^{\perp}(\tilde{\sigma}_m), \hat{W}^T_m\Delta \mu_v \leq 0 & \text{if} \ y < y_m \end{array} \right., \nonumber \\
& \Delta\hat{W}_{m,c} = \left\{ \begin{array}{cc} \sqrt{n_s}\Delta \mu_v & \text{if} \ y > y_m \\ -\sqrt{n_s}\Delta \mu_v & \text{if} \ y < y_m \end{array} \right.\nonumber 
\end{align}
We will specify below how $\tilde{\rho}$ is picked. Given these definitions and Assumption \ref{ass:memory} (point (ii)) we have that
\begin{equation}
M_r(\mu_c, k_\mu, \hat{S}_m, \tilde{x}) = M_r(\mu, k_\mu, \hat{S}_m, \tilde{x}) + \Delta \mu_v \nonumber 
\end{equation}
Therefore
\begin{align}
& \omega_{f,o}(z) = \norm{e_m} = \norm{y - y_m} = \norm{y - \hat{W}^T_m(\hat{\sigma}_m+M_r)}, \nonumber \\
& \omega_{f,o}(z_c) = \norm{e_{m,c}} = \norm{y_c - y_{m,c}} \nonumber \\
& = \norm{y - \hat{W}^T_{m,c}(\hat{\sigma}_m+M_r+\Delta \mu_v)} \nonumber \\
& = \norm{y - (\hat{W}_{m}+\Delta\hat{W}_{m,c})^T(\tilde{\sigma}_m + \Delta \mu_v)} \nonumber \\
&= \norm{y - \hat{W}^T_{m}(\tilde{\sigma}_m + \Delta \mu_v) - \Delta\hat{W}^T_{m,c}(\tilde{\sigma}_m + \Delta \mu_v)} \nonumber 
\end{align}
Now, we prove on a case by case basis. \\
{\em (i) Case $y > y_m$}: In this case, from the definition of $\Delta \mu_v$ and $\Delta\hat{W}_{m,c}$ it follows that
\begin{align}
\hat{W}^T_{m}\Delta \mu_v \geq 0, \ \Delta\hat{W}^T_{m,c}\tilde{\sigma}_m = 0, \ \Delta\hat{W}^T_{m,c}\Delta \mu_v = \frac{\tilde{\rho}^2}{2\sqrt{n_s}} \nonumber 
\end{align}
Using these observations we get that
\begin{align}
& e_{m,c} = y - y_m -\hat{W}^T_{m}\Delta \mu_v -\Delta\hat{W}^T_{m,c}(\tilde{\sigma}_m + \Delta \mu_v) \nonumber \\
& \leq y - y_m -\Delta\hat{W}^T_{m,c}\Delta \mu_v = y -y_m -\frac{\tilde{\rho}^2}{2\sqrt{n_s}}\nonumber
\end{align}
Now, pick $\tilde{\rho} (\leq \rho)$ such that $e_{m,c} \geq 0$. We can pick such a $\tilde{\rho}$ by picking it to be sufficiently small. Hence, for such a $\tilde{\rho}$ and given that $e_m > 0$ and $e_{m,c} \geq 0$, we have that 
\begin{align}
& \norm{e_{m,c}} \leq \norm{y - y_m -\Delta\hat{W}^T_{m,c}\Delta \mu_v} \nonumber \\
& = \norm{e_m -\Delta\hat{W}^T_{m,c}\Delta \mu_v} = \norm{e_m} -\frac{\tilde{\rho}^2}{2\sqrt{n_s}} \nonumber
\end{align} 
Now, by definition $\norm{z - z_c} = \tilde{\rho}$. Set $\epsilon^{*} = \frac{\rho^2}{2\sqrt{n_s}}$. So in this case, for each $c \in [0,c_{f,i}]$, if $\omega_{f,o}(z) = \norm{e_m} \leq c + \epsilon^{*}$, we can pick an appropriate $\tilde{\rho} \leq \rho$ and $z_c$ as defined above such that $\norm{z - z_c} = \tilde{\rho} \leq \rho$ and 
$\omega_{f,o}(z_c) = \norm{e_{m,c}} \leq c$.
{\em (ii) Case $y < y_m$}:  In this case, from the definition of $\Delta \mu_v$ and $\Delta\hat{W}_{m,c}$ it follows that
\begin{equation}
\hat{W}^T_{m}\Delta \mu_v \leq 0, \ \Delta\hat{W}^T_{m,c}\tilde{\sigma}_m = 0, \ \Delta\hat{W}^T_{m,c}\Delta \mu_v = -\frac{\tilde{\rho}^2}{2\sqrt{n_s}} \nonumber 
\end{equation}
Using these observations we get that
\begin{align}
& e_{m,c} = y - y_m -\hat{W}^T_{m}\Delta \mu_v -\Delta\hat{W}^T_{m,c}\Delta \mu_v \nonumber \\
& \geq y - y_m -\Delta\hat{W}^T_{m,c}\Delta \mu_v = y - y_m + \frac{\tilde{\rho}^2}{2\sqrt{n_s}} \nonumber
\end{align}
Now, pick $\tilde{\rho} (\leq \rho)$ such that $e_{m,c} \leq 0$. We can pick such a $\tilde{\rho}$ by picking it to be sufficiently small. Hence, for such a $\tilde{\rho}$ and given that $e_m < 0$ and $e_{m,c} \leq 0$, we have that 
\begin{equation}
\norm{e_{m,c}} \leq \norm{y - y_m + \frac{\tilde{\rho}^2}{2\sqrt{n_s}}} = \norm{e_m + \frac{\tilde{\rho}^2}{2\sqrt{n_s}}} = \norm{e_m} -\frac{\tilde{\rho}^2}{2\sqrt{n_s}} \nonumber
\end{equation} 
Now, by definition $\norm{z - z_c} = \tilde{\rho}$. Set $\epsilon^{*} = \frac{\rho^2}{2\sqrt{n_s}}$. So in this case, for each $c \in [0,c_{f,i}]$, if $\omega_{f,o}(z) = \norm{e_m} \leq c + \epsilon^{*}$, we can pick an appropriate $\tilde{\rho} \leq \rho$ and $z_c$ as defined above such that $\norm{z - z_c} = \tilde{\rho} \leq \rho$ and 
$\omega_{f,o}(z_c) = \norm{e_{m,c}} \leq c$.
This completes the proof of statement s1.

Now, statement s1, Assumption \ref{ass:memory}, Assumption \ref{ass:disturbance}, Assumption \ref{ass:bdrylayer-average-dynamics}, Assumption \ref{ass:initialset}, and Assumption \ref{ass:continuity} together satisfy Assumption 2, Assumption 3, Assumption 4, Assumption 7 and Assumption 8 of \cite{teel2003unified}. Then the following statement follows from Claim 2 of Appendix A in \cite{teel2003unified}: (statement s2) $\exists \ T^{*} > 0$ so that for each $T \geq T^{*}$ and $\tilde{\delta} > 0$, there exists $\epsilon^{*} > 0$ such that $x_s(t) \in \mathcal{X}_s+\tilde{\eta}\mathcal{B}$ (some $\tilde{\eta} \leq \eta/2$), $z(0) = \mathcal{K}_f$, $\epsilon \in (0, \epsilon^{*}]$, imply that the solution $z(.)$ of the overall dynamics given by Eq. \eqref{eq:estimator-dynamics}, the solution $z_{bl}(.)$ of the boundary layer dynamics (Definition \ref{def:bdrylayerdyn}) with $z_{bl}(0) = z(0)$ satisfy
\begin{align}
& \vert \omega_{f,o}(z(t)) - \omega_{f,o}(z_{bl}(t))\vert \leq \tilde{\delta}, \ \forall \ t \in [0, T], \nonumber \\
& z(t) \in \mathcal{K}_f, \ \forall \ t \in [T^{*}, T] \nonumber 
\end{align}
And from claim 5 of Appendix A in \cite{teel2003unified}: (statement s3) for each $\tilde{\delta} > 0$, $T > 0$, there exists $\epsilon^{*} > 0$ such that $\epsilon \in (0, \epsilon^{*}]$, $x_s(0) = x^{av}_s(0) \in \mathcal{K}_s$, $z(0) \in \mathcal{K}_f$ imply that $x_s(t) \in \mathcal{X}_s + \tilde{\eta}\mathcal{B} \ \forall \ t \in [0,T^{*}/\epsilon]$, for some $\tilde{\eta} \leq \eta/2$.

Let $\tilde{\epsilon}^{*}$ be the minimum of the $\epsilon^{*}$ drawn from statement s2 and statement s3 for $(T, \tilde{\delta} = \alpha(\overline{\delta}+\delta_m))$, where $T \geq T^{*}$ and $T^{*}$ is drawn from statement s2. Consider an arbitrary $\epsilon \in (0, \tilde{\epsilon}^{*}]$, such that $\epsilon < 1$. Let $z(.)$ be the solution of the overall dynamics given by Eq. \eqref{eq:estimator-dynamics} starting from $z(0) \in \mathcal{K}_f$, and $z_{bl}$ be the corresponding solution of the boundary layer dynamics starting from $z_{bl}(0) = z(0)$. Then, combining statement s2 and s3 we get that
\begin{equation}
\omega_{f,o}(z(t)) \leq \omega_{f,o}(z_{bl}(t)) + \alpha(\overline{\delta}+\delta_m), \forall \ t \in [0,T] \nonumber
\end{equation}
Since $T \geq T^{*}$, $z(T) \in \mathcal{K}_f$. Now, repeatedly applying statement s2 and statement s3, we get that
\begin{equation}
\omega_{f,o}(z(t)) \leq \omega_{f,o}(z_{bl}(t)) + \alpha(\overline{\delta}+\delta_m), \ \forall \ t 
\label{eq:omegafo-bound}
\end{equation}
Now, by the definition of the boundary layer dynamics
\begin{equation}
\omega_{f,o}(z_{bl}(t)) = \norm{W^T\sigma(t) + \delta_t - \hat{W}^T_m(0)(\hat{\sigma}_m(0)+\alpha M^{bl}_r)} \nonumber 
\end{equation}
That is
\begin{align}
& \omega_{f,o}(z_{bl}(t)) = \norm{W^T\sigma(t) +\delta_t - \hat{W}^T_m(0)(\hat{\sigma}_m(0)+\alpha M^{bl}_r)} \nonumber \\
& = \norm{W^T\sigma(t) + \delta_t - \hat{W}^T_m(0)(\hat{\sigma}_m(0)+\alpha\sigma(t)+\alpha\delta_\sigma)}, \nonumber 
\end{align}
where $\norm{\delta_\sigma} \leq \delta_m$. Given the initial condition for the dynamics Eq. \eqref{eq:estimator-dynamics} and Eq. \eqref{eq:estimator-dynamics-1}, using Lemma \ref{lem:est-mem-feas-sol} we have that
\begin{align}
& \omega_{f,o}(z_{bl}(t)) \nonumber \\
& = \norm{W^T\sigma(t) +\delta_t - \frac{W^T}{(1+\alpha)}(\hat{\sigma}(0)+\alpha\sigma(t)+\alpha\delta_\sigma)} \nonumber
\end{align}
That is 
\begin{align}
& \omega_{f,o}(z_{bl}(t)) \leq \frac{\norm{W^T(\sigma(t)-\hat{\sigma}(0))+\delta_t}}{1+\alpha} + O(\alpha(\overline{\delta}+\delta_m)) \nonumber \\
& =  \frac{\norm{e^{bl}(t)}}{1+\alpha} + O(\alpha(\overline{\delta}+\delta_m)) \leq \frac{\vert e^{bl,max}\vert }{1+\alpha} + O( \alpha(\overline{\delta}+\delta_m)). \nonumber 
\end{align}
Combining this with Eq. \eqref{eq:omegafo-bound} we get the final result. $\blacksquare$

\section*{Appendix C: Proof of Theorem \ref{thm:stability-1}}
We first prove the following Lemmas.

\begin{lemma}
{\it Let $\tilde{P}$ be the matrix solution to the continuous time algebraic Ricatti equation (CARE),
\beq A^T\tilde{P} + \tilde{P}A - 1/K_r\tilde{P}BB^T\tilde{P} = -Q = -K_vI \eeq 
that stabilizes the pair $(A,B)$. If $\lambda_i$ denotes the $i$th eigenvalue of $\tilde{P}$, then
\beq O(K_v) \geq \lambda_i \geq O(\sqrt{K_v}). \eeq}
\label{lem:eigtP}
\end{lemma}
The proof of this lemma follows from Theorem $1$ in \cite{kwon1996bounds}.

\begin{lemma}
{\it The matrix $P$ for the MRAC controller in section \ref{sec:controllaw} satisfies $\norm{PB}_F \leq O(K_v)^{3/4}$.}
\label{lem:normP}
\end{lemma}
{\it Proof}: Consider the Lyapunov equation, $A^T_{\text{ref}}P + PA_{\text{ref}} = -Q$, where $A_{\text{ref}} = A - 1/K_rBB^T\tilde{P}$ and $\tilde{P}$ is the matrix solution defined in Lemma \ref{lem:eigtP}. Substituting for $A^T_{\text{ref}}$, we get that
\beq 1/K_r\tilde{P}BB^TP +1/K_rPBB^T\tilde{P} -(A^TP + PA) = Q. \eeq

Because $A_{\text{ref}}$ is Hurwitz, the above equation has a unique matrix solution that is positive-definite and symmetric. Multiplying on the left by $\tilde{P}^{-1}$, we get that
\beq BB^TP + \tilde{P}^{-1}PBB^T\tilde{P} - K_r\tilde{P}^{-1}(A^TP + PA) = K_r\tilde{P}^{-1}Q. \nonumber \eeq

Taking trace on both sides, we get that
\beq
\frac{2}{K_r}\text{Tr}\{BB^TP\} = \left(\text{Tr}\{\tilde{P}^{-1}Q\} +\text{Tr}\{\tilde{P}^{-1}(A^TP + PA)\} \right). \nonumber
\eeq

From Theorem $3$ in \cite{mrabti1992bounds}, it trivially follows that $\tilde{s}_1 \leq O(K_v)$, where $\tilde{s}_1$ is the maximum eigenvalue of $P$. Using this observation and Lemma \ref{lem:eigtP}, we get that 
\beq \frac{2}{K_r}\text{Tr}\{B^TPB\} \leq O(\sqrt{K_v}). \eeq 

That is,
\beq \frac{2}{K_r}\text{Tr}\{B^TP^{\frac{1}{2}}P^{\frac{1}{2}}B\} \leq O(\sqrt{K_v}). \eeq 

That is,
\beq \frac{2}{K_r}\norm{P^{\frac{1}{2}}B}^2_F \leq O(\sqrt{K_v}), \text{Or} \  \norm{P^{\frac{1}{2}}B}_F \leq O(K_v)^{1/4}. \eeq

That is,
\beq \norm{PB}_F \leq \norm{P^{1/2}}_F \norm{P^{\frac{1}{2}}B}_F \leq O(K_v)^{3/4} \blacksquare \nonumber \eeq

We now prove theorem \ref{thm:stability-1}. {\it Proof}:

The functions on the right hand side of the equations of the closed loop system (Eq. \eqref{eq:memop}, Eq. \eqref{eq:NNupdate}, Eq. \eqref{eq:claw}, Eq. \eqref{eq:sys-lmu}) are all continuous with respect to the variables $x, \hat{W}, \hat{V}, \mu$. Second, for each $\{x, \hat{W}, \hat{V}, \mu\}$, the functions on the right-hand side of the closed loop system (Eq. \eqref{eq:memop}, Eq. \eqref{eq:NNupdate}, Eq. \eqref{eq:sys-lmu}) are all measurable w.r.t $t$ because (i) they are piece-wise constant functions for a given $\{x, \hat{W}, \hat{V}, \mu\}$ with discontinuities at finite no. of points and (ii) such functions are Lebesgue measurable. Third, for each $\{x, \hat{W}, \hat{V}, \mu\}$, and a finite radius ball around this point the functions on the right hand side are bounded. Then using Proposition S1 of \cite{cortes2009discontinuous} it follows that the Caratheodory solution of the closed loop system exist. We introduce a function $L_e$ by, where $L_e := \frac{1}{2}e^TPe$. Consider the following positive definition function:
\beq 
L  =  L_e + \frac{1}{2\gamma_w} \text{Tr}\{\tilde{W}\tilde{W}^T\} + \frac{1}{2\gamma_v} \text{Tr}\{\tilde{V}\tilde{V}^T\}. 
\eeq 

Let $Q = K_v I$, where $K_v > 0$ and $I$ is the identity matrix and $Q$ is the same matrix used to derive the LQR controller gain. Let $P$ be the positive definite matrix solution to the Lyapunov equation,
\beq A_{\text{ref}}^TP + PA_{\text{ref}} = -2Q \label{eq:Plyapeq}\eeq 


Since, in between the abrupt changes, $f_t, W_t, V_t$ are constants, we drop the subscript $t$ in the following part of the analysis for the period between the abrupt changes. For the MRAC controller, the LQR control gain $K_{\text{lqr}}$ in $u_{bl} = -K_{\text{lqr}}x$ is such that $A - BK_{\text{lqr}} = A_{\text{ref}}$.  Hence, for the Caratheodory solution, between the abrupt changes
\[\dot{e} = \dot{x} - \dot{x}_{\text{ref}} = A_{\text{ref}}e + B(v + \tilde{f}),\]

where $\tilde{f} = f - \hat{f}$. Therefore, between the abrupt changes
\begin{align} \dot{L}_e & = \frac{1}{2}\dot{e}^TPe + \frac{1}{2}e^TP\dot{e} \nonumber \\
& =  \frac{1}{2}\left(e^TA^T_{\text{ref}}Pe + e^TPA_{\text{ref}}e\right) \nonumber \\
& + \frac{1}{2}\left((v + \tilde{f})^TB^TPe + e^TPB(v + \tilde{f})\right) 
\end{align}

Therefore, between the abrupt changes, the time derivative of $L$ for the Caratheodory solution is given by
\begin{align} 
\dot{L} = & \dot{L}_e + \frac{1}{2\gamma_w} \text{Tr}\{\dot{\tilde{W}}\tilde{W}^T\}+ \frac{1}{2\gamma_w} \text{Tr}\{\tilde{W}\dot{\tilde{W}}^T\}  \nonumber\\
& + \frac{1}{2\gamma_v} \text{Tr}\{\dot{\tilde{V}}\tilde{V}^T\} + \frac{1}{2\gamma_v} \text{Tr}\{\tilde{V}\dot{\tilde{V}}^T\}.
\end{align}

Using the trace identity $\text{Tr}\{A^TB\} = \text{Tr}\{B^TA\}$, we can simplify the above expression as
\beq
\dot{L} = \dot{L}_e + \frac{1}{\gamma_w}\text{Tr}\{\tilde{W}^T\dot{\tilde{W}}\} + \frac{1}{\gamma_v}\text{Tr}\{\tilde{V}^T\dot{\tilde{V}}\} \eeq

Substituting for $\dot{L}_e$ from above, we get that
\begin{align} 
& \dot{L} = -e^TQe + e^TP(x)R(v + \tilde{f}) \nonumber\\
& + \frac{1}{\gamma_w}\text{Tr}\{\tilde{W}^T\dot{\tilde{W}}\} + \frac{1}{\gamma_v}\text{Tr}\{\tilde{V}^T\dot{\tilde{V}}\}.
\label{eq:derL} 
\end{align}

Define $Q_d$ to be the bound on the norm of the state trajectory of the reference model or the desired trajectory and its derivatives up to second order for each of the controllers respectively. Denote the compact subset of $\mathbb{R}^n$ within which the $\overline{\delta}$-approximation holds by $\tilde{\mathcal{U}}_f$. Let this set be given by $\tilde{\mathcal{U}}_f = \{\tilde{x} \big\vert \norm{\tilde{x}}_2 \leq \tilde{r}_f\}$. We can show trivially that there exists constants $d_1$ and $d_2$ such that
\beq \tilde{x} \leq d_1 Q_d + d_2 \norm{e}_2. \eeq

We introduce the following definitions: $r_f := \frac{\tilde{r}_e - d_1Q_d}{d_2}, \ \mathcal{U}_f := \{e \big\vert \norm{e}_2 \leq r_f\}$. It follows that, when $\norm{e}_2 \leq r_f$, $\norm{\tilde{x}}_2 \leq \tilde{r}_f$, implying that the $\overline{\delta}$-approximation holds when $e \in \mathcal{U}_f$. Hence,
\[ f(\tilde{x}) = W^T\sigma(V^Tx_e) + \overline{\delta} \ \forall \ \tilde{x} \in \tilde{\mathcal{U}}_f.\]
Hence,
\[\tilde{f} =  W^T\sigma(V^Tx_e) -\hat{W}^T\sigma(\hat{V}^Tx_e) -\hat{W}^TM_r + \overline{\delta} \ \forall \ \tilde{x} \in \tilde{\mathcal{U}}_f. \]

Adding and subtracting $W^T\sigma(\hat{V}^Tx_e)$ to $\tilde{f}$, we get that
\begin{align}
\tilde{f} & = W^T\sigma(\hat{V}^Tx_e) -\hat{W}^T\sigma(\hat{V}^Tx_e) \nonumber \\
& +W^T( \sigma(V^Tx_e) -\sigma(\hat{V}^Tx_e)) -\hat{W}^TM_r +  \overline{\delta}.
\end{align}

Combining the first two terms we get that
\[ \tilde{f} = \tilde{W}^T\sigma(\hat{V}^Tx_e) +W^T( \sigma(V^Tx_e) -\sigma(\hat{V}^Tx_e)) -\hat{W}^TM_r +  \overline{\delta}.\]

Using Taylor's series expansion for the second term we get that
\[ \tilde{f} = \tilde{W}^T\sigma(\hat{V}^Tx_e) +W^T\hat{\sigma}^{'}\tilde{V}^Tx_e +W^TO(\tilde{V}^Tx_e)^2 -\hat{W}^TM_r +  \overline{\delta}. \]

Adding and subtracting $\hat{W}^T\hat{\sigma}^{'}\tilde{V}^Tx_e$, and rearranging terms, we get that
\[ \tilde{f} = \tilde{W}^T\left(\hat{\sigma} -\hat{\sigma}^{'}\hat{V}^Tx_e\right) +\hat{W}^T \hat{\sigma}^{'}\tilde{V}^Tx_e -\hat{W}^TM_r +w_1,\]
where $w_1 = \tilde{W}^T\hat{\sigma}^{'}V^Tx_e + W^TO(\tilde{V}^Tx_e)^2  + \overline{\delta}$. Define, $q_\mu := e^TP(x)B$. Then substituting for $\tilde{f}$ in \eqref{eq:derL} we get that
\begin{align} 
& \dot{L} = -e^TQe + q_\mu(v + w_1- \hat{W}^TM_r)  \nonumber\\
 &  +\frac{1}{\gamma_w} \text{Tr}\{\tilde{W}^T\dot{\tilde{W}}\} + q_\mu\tilde{W}^T\left(\hat{\sigma} - \hat{\sigma}^{'}\hat{V}^Tx_e\right) + \frac{1}{\gamma_v}\text{Tr}\{\tilde{V}^T\dot{\tilde{V}}\} \nonumber\\
 & + q_\mu\hat{W}^T \hat{\sigma}^{'}\tilde{V}^Tx_e.
\end{align}

Using the identity $\text{Tr}\{AB\} = \text{Tr}\{BA\}$ we get that
\begin{align} 
& \dot{L} = -e^TQe + q_\mu(v + w_1- \hat{W}^TM_r) \nonumber\\
 &  + \frac{1}{\gamma_w}\text{Tr}\{\tilde{W}^T\dot{\tilde{W}} + \gamma_w\tilde{W}^T\left(\hat{\sigma} - \hat{\sigma}^{'}\hat{V}^Tx_e\right)q_\mu\} \nonumber\\
 & + \frac{1}{\gamma_v}\text{Tr}\{\tilde{V}^T\dot{\tilde{V}} + \gamma_v\tilde{V}^Tx_eq_\mu\hat{W}^T \hat{\sigma}^{'}\}.
\end{align}

It follows from the NN update laws \eqref{eq:NNupdate} that the last two terms vanish and two new terms given by $\kappa\norm{e}_2\text{Tr}\{\tilde{W}^T(W-\tilde{W})\}$ and $\kappa\norm{e}_2\text{Tr}\{\tilde{V}^T(V-\tilde{V})\}$ gets added. Hence, the expression simplifies as
\begin{align} 
\dot{L} & = -e^TQe + q_\mu(v + w_1 - \hat{W}^TM_r) \nonumber \\
& + \kappa\norm{e}_2\text{Tr}\{\tilde{W}^T(W-\tilde{W})\} + \kappa\norm{e}_2\text{Tr}\{\tilde{V}^T(V-\tilde{V})\}. \nonumber
\end{align}


Substituting for $v$ from Eq. \eqref{eq:robterm} we get that
\begin{align} 
\dot{L} & = -e^TQe + q_\mu w_1 - q_\mu \hat{W}^TM_r \nonumber\\
& - \norm{q_\mu}_2 k_z(\norm{\hat{W}}_F + \norm{\hat{V}}_F + Z_m) \norm{e}_2 \nonumber\\
& + \kappa\norm{e}_2\text{Tr}\{\tilde{W}^T(W-\tilde{W})\} + \kappa\norm{e}_2\text{Tr}\{\tilde{V}^T(V-\tilde{V})\}. \nonumber
\end{align}

Using the fact that $e^TQe = K_ve^Te$ we get
\begin{align} 
\dot{L} & \leq -K_ve^Te + q_\mu w_1 - q_\mu \hat{W}^TM_r  \nonumber\\
&  - \norm{q_\mu}_2 k_z(\norm{\hat{W}}_F + \norm{\hat{V}}_F  + Z_m) \norm{e}_2 \nonumber\\
& + \kappa\norm{e}_2\text{Tr}\{\tilde{W}^T(W-\tilde{W})\} + \kappa\norm{e}_2\text{Tr}\{\tilde{V}^T(V-\tilde{V})\}. \nonumber
\label{eq:part1eq2}
\end{align}

From Lemma  \ref{lem:w1} we get that
\beq
\norm{w_1}_2 \leq b_1+ b_2\norm{\tilde{Z}}_F + c_2Z_m\norm{e}_2 + c_{3}\norm{\hat{Z}}_F\norm{e}_2. 
\eeq

Using the above inequality and rearranging terms we get that
\begin{align} 
\dot{L} & \leq -K_v\norm{e}_2^2 - q_\mu\hat{W}^T M_r \nonumber\\
&  - \norm{q_\mu}_2 Z_m \norm{e}_2(k_z - c_2) - \norm{q_\mu}_2 \norm{\hat{Z}}_F \norm{e}_2 (k_z - c_3)\nonumber\\
& \norm{q_{\mu}}_2\left(b_1+ b_2\norm{\tilde{Z}}_F\right) + \kappa\norm{e}_2\text{Tr}\{\tilde{Z}^T(Z-\tilde{Z})\}.
\label{eq:mucondn}
\end{align}

Since $k_z \geq \max\{c_2,c_3\}$,
\begin{align}
\dot{L} \leq & -K_v\norm{e}_2^2 + \norm{q_{\mu}}_2\left(b_1+ b_2\norm{\tilde{Z}}_F\right) -q_\mu\hat{W} M_r \nonumber\\
& + \kappa\norm{e}_2\text{Tr}\{\tilde{Z}^T(Z-\tilde{Z})\}.
\end{align}

For now, lets assume that $\norm{\mu}_F \leq \overline{\mu}$. Then the above expression can be simplified as
\begin{align}
\dot{L} \leq & -K_v\norm{e}_2^2 + \norm{q_{\mu}}_2\left(\tilde{b}_1+ \tilde{b}_2\norm{\tilde{Z}}_F\right) \nonumber\\
& + \kappa\norm{e}_2\text{Tr}\{\tilde{Z}^T(Z-\tilde{Z})\},
\end{align}
where both $\tilde{b}_1$ and $\tilde{b}_2$ are constants that depend on $\overline{\mu}$. It follows that there exist constants $\tilde{b}_3 \leq O(K_v)^{3/4}$ and $\tilde{b}_4 \leq O(K_v)^{3/4}$ such that
\begin{align}
\dot{L} & \leq -\norm{e}_2 \left(K_v \norm{e}_2 - \tilde{b}_3 - \tilde{b}_4\norm{\tilde{Z}}_F\right) \nonumber \\
& -\norm{e}_2 \left( \kappa\norm{\tilde{Z}}_F\left(\norm{\tilde{Z}}_F - Z_m\right) \right).
\end{align}

We introduce $\tilde{b}_5$, where $\tilde{b}_5 := \frac{\kappa Z_m + \tilde{b}_4}{2 \kappa}$. Completing squares, we get the following:
\begin{equation}
\dot{L} \leq -\norm{e}_2 \left(K_v\norm{e}_2 - \tilde{b}_3 + \kappa(\norm{\tilde{Z}}_F - \tilde{b}_5)^2 - \kappa \tilde{b}^2_5\right).
\end{equation}

Hence, $\dot{L} < 0$ when either
\beq \norm{e}_2 > \frac{\tilde{b}_3 + \kappa \tilde{b}^2_5}{K_v} = r_e,  \text{Or} \ \norm{\tilde{Z}}_F > \tilde{b}_5 + \sqrt{\tilde{b}^2_5 + \frac{\tilde{b}_3}{\kappa}}  = r_z/2. \label{eq:cond3} \eeq 


Therefore, for the Caratheodory solution, $\dot{L}$ is strictly negative ($ < 0$, except at instants of abrupt jumps which are of measure zero) outside a compact set defined by the radii $r_e$ and $r_z$. Hence, it follows that the signals $e$ and $\tilde{Z}$ of the Caratheodory solution are uniformly bounded. To ensure that the $\overline{\delta}$-approximation is valid within the compact set defined by the radii $r_e$, $\mathcal{U}_e = \{e \big \vert \norm{e}_2 \leq r_e \}$, it should be a strict subset of $\mathcal{U}_f$. Since $\kappa = K_v^{3/4}$, $\kappa \tilde{b}^2_5 = O(\kappa +\tilde{b}_4 +\tilde{b}^2_4/\kappa) = O(K_v)^{3/4}$. Consequently, the numerator $\tilde{b}_3 + \kappa \tilde{b}^2_5$ in \eqref{eq:cond3} is $= O(K_v)^{3/4}$. Hence, we can choose the gain $K_v$ to be large enough that the compact set defined by $\mathcal{U}_e = \{e \big \vert \norm{e}_2 \leq r_e \}$ is a strict subset of the compact set $\mathcal{U}_f$. To ensure that the transients of the Caratheodory solution don't overshoot the set $\mathcal{U}_f$ the gain $K_v$ is set large enough that $\lambda_{max}(P)\gamma^{0.25}_vr^2_e/\lambda_{min}(P) +r^2_z/(4\gamma^{0.75}_v\lambda_{min}(P)) +2Z^2_m/(\gamma^{0.75}_v\lambda_{min}(P))  = \bar{r}^2 < r^2_f$, where 
the last term on the left accounts for abrupt changes given by $\norm{\Delta Z}_F \leq Z_m$. This will ensure that $e$ of the Caratheodory solution stays within $r_f$ all the time when the signals $e$ and $\tilde{Z}$ start from within the radii $r_e$ and $r_z/2$ respectively. Earlier, we assumed that $\norm{\mu}_F$ is bounded. From the memory update equations \eqref{eq:memop} it follows trivially that $\mu$ of the Caratheodory solution is bounded when $\tilde{Z}$ and $e$ are bounded. Denote this bound by $\tilde{\mu}$ when $\norm{e}_2 \leq \bar{r}$ and $\norm{\tilde{Z}}_F \leq \sqrt{\gamma_v}\bar{r}$. We can set $\overline{\mu} > \tilde{\mu}$ to ensure consistency of the bound used in the derivation above. This completes the proof. $\blacksquare$

\begin{IEEEbiography}
{Deepan Muthirayan}
is currently a Post-doctoral Researcher in the department of Electrical Engineering and Computer Science at University of California at Irvine. He obtained his Phd from the University of California at Berkeley (2016) and B.Tech/M.tech degree from the Indian Institute of Technology Madras (2010). His doctoral thesis work focused on market mechanisms for integrating demand flexibility in energy systems. Before his term at UC Irvine he was a post-doctoral associate at Cornell University where his work focused on online scheduling algorithms for managing demand flexibility. His current research interests include control theory, machine learning, learning for control, online learning, online algorithms, game theory, and their application to smart systems.
\end{IEEEbiography}

\begin{IEEEbiography}
{Pramod Khargonekar}

received B. Tech. Degree in electrical engineering in 1977 from the Indian Institute of Technology, Bombay, India, and M.S. degree in mathematics in 1980 and Ph.D. degree in electrical engineering in 1981 from the University of Florida, respectively. He was Chairman of the Department of Electrical Engineering and Computer Science from 1997 to 2001 and also held the position of Claude E. Shannon Professor of Engineering Science at The University of Michigan.  From 2001 to 2009, he was Dean of the College of Engineering and Eckis Professor of Electrical and Computer Engineering at the University of Florida till 2016. After serving briefly as Deputy Director of Technology at ARPA-E in 2012-13, he was appointed by the National Science Foundation (NSF) to serve as Assistant Director for the Directorate of Engineering (ENG) in March 2013, a position he held till June 2016. Currently, he is Vice Chancellor for Research and Distinguished Professor of Electrical Engineering and Computer Science at the University of California, Irvine. His research and teaching interests are centered on theory and applications of systems and control. He has received numerous honors and awards including IEEE Control Systems Award, IEEE Baker Prize, IEEE CSS Axelby Award, NSF Presidential Young Investigator Award, AACC Eckman Award, and is a Fellow of IEEE, IFAC, and AAAS.

\end{IEEEbiography}

\end{document}